\documentclass[12pt]{iopart}
\usepackage{epsfig,bm}
\usepackage{graphicx}

\newcommand{\bea}{\begin{eqnarray}}
\newcommand{\eea}{\end{eqnarray}}
\newcommand{\nn}{\nonumber}
\newcommand{\be}{\begin{equation}}
\newcommand{\ee}{\end{equation}}
\newcommand{\bd}{\begin{displaymath}}
\newcommand{\ed}{\end{displaymath}}

\newcommand{\bra}{\langle}
\newcommand{\ket}{\rangle}

\newcommand{\Prob}{{\rm Prob}}
\newcommand{\erf}{{\rm erf}}

\newcommand{\vn}{{\bm n}}

\newcommand{\cN}{\ensuremath{\mathcal{N}}}
\newcommand{\cS}{{\cal S}}

\usepackage{epsfig}
\usepackage{graphicx}
\usepackage{eufrak}

\newcommand{\bphi}{{\mbox{\boldmath $\phi$}}}

\begin{document}

\title{Effects of Economic Interactions on Credit Risk}
\author{JPL Hatchett and R K\"uhn }
\address{Laboratory for mathematical neuroscience, RIKEN BSI, Hirosawa
  2-1, Saitama 351-0198, Japan}
\address{Department of Mathematics, King's College London, The Strand,
London WC2R 2LS, United Kingdom}

\begin{abstract}
We study a credit risk model which captures effects of economic
interactions on a firm's default probability. Economic interactions
are represented as a functionally defined graph, and the existence
of both cooperative, and competitive, business relations is taken
into account. We provide an analytic solution of the model in a
limit where the number of business relations of each company is
large, but the overall fraction of the economy with which a given
company interacts may be small. While the effects of economic
interactions are relatively weak in typical (most probable)
scenarios, they are pronounced in situations of economic stress, and
thus lead to a substantial fattening of the tails of loss
distributions in large loan portfolios. This manifests itself in
a pronounced enhancement of the Value at Risk computed for interacting
economies in comparison with their non-interacting counterparts.\\~\\
Date:~~December 6, 2005
\end{abstract}


\ead{\mailto{ hatchett@brain.riken.jp}, \mailto{kuehn@mth.klc.ac.uk} }

\section{Introduction}

The proper quantification of credit risk poses a complex mix of
problems, as important credit risk parameters  such as default
rates, recovery rates or exposures, fluctuate substantially in time
even on a high portfolio aggregation level \cite{Keenan00}. This
results in large unexpected losses in loan portfolios, for which
banks are required to hold equity capital as a loss buffer. To
determine the appropriate level of equity capital for banks' loan
portfolios is one main focus of the regulatory consultive process
known as Basel II \cite{BISII}. Accordingly, credit risk modelling
has been a focus of intense research in recent years \cite{CM,
CRisk,  Kealhofer98, Davis00, Lando98, Duffie98, Jarrow01, Rogge02,
Jarrow03, Das05, Yu03, Duffie99, Giesecke03, Weber02, Tasche03,
Eglo03, Gordy00, Gordy01, NeuKuehn04}, although considering the risk
premium when pricing interest rates goes back some time
\cite{Merton74}.

The assessment of credit ratings by assessment agencies such as
Moody's and S$\&$P allow some statistical assessment of the credit
quality of individual offerings or particular companies. However, it
is clearly essential when considering the risk of a basket of loans
that the correlations between the members of the portfolio are taken
into account. One systematic approach is to replace the number of
firms in a portfolio with an {\em effective} number of independent
firms \cite{Moodys97}. By boosting the contribution of each firm to
keep the mean loss constant, this introduces a larger variance of
losses, in at attempt to capture the risk caused by correlations
between firms. JPMorgan's CreditMetrics approach \cite{CM} (see also
Credit Suisse Financial Products's CreditRisk$^{+}$ \cite{CRisk} and
\cite{Gordy00} for a detailed comparison of the two) tries to model
the correlations between firms in credit quality using the
observable correlations in equity value of the firms. An intuitively
appealing approach is to assume that the default intensity depends
on some set of macroscopic economic factors (e.g. interest rates,
growth rates, oil prices etc.), the so-called reduced form model
\cite{Lando98, Duffie98}. Thus the default rates of different firms
are coupled via some limited number of factors, but given the
factors the default rates are independent. In structural models 
\cite{Merton74, Algo99}, the dependence on macro-economic factors is 
understood in terms of correlations in the dynamics of asset returns 
of different companies, leading to correlations in default rates via 
correlated dynamics of returns. More
involved approaches have modelled interactions between firms in the
wider economy by introducing changes to a firm's default intensity
upon the default of another firm \cite{Davis00}, or via a copula
function \cite{Rogge02}. The quantification of these correlations
via simulations is discussed in \cite{Duffie99}.

The main purpose of the present contribution is specifically to
expand upon recent modelling and analytic descriptions of the
influence of counter-party risk \cite{Davis00, Jarrow01,Jarrow03,
Yu03, Giesecke03, Weber02, Tasche03, Eglo03, NeuKuehn04}.
Counter-party risk addresses the fact that a given firm's economic
health is strongly influenced by the performance within the network
of companies with which it has direct economic interactions.
Economic interactions are here to be read in a broad sense as any
form of relation which is likely to mutually influence performance,
though not necessarily in symmetric ways. Specifically, a defaulting
firm within this network of counter-parties will affect a company's
own default probability
--- reducing it, if the defaulting firm was a competitor, or
increasing it, if the relation was of a cooperative nature. When the
default probability is increased, this process is known as {\em
credit contagion} and has been considered in e.g. \cite{Weber02,
Eglo03} while the incorporation of counter-party risk into a reduced
form model was introduced in \cite{Jarrow01}. In what follows we
consider the dynamics of individual firm defaults and their
influence on loss distributions. More subtle effects such as credit
quality migration are, as yet, not taken into account.

The importance of direct functional interactions in the analysis of
risk is not restricted to credit risk. In fact the role of
interactions is much more obvious in the context of operational
risk, where sequential, functionally induced failures of mutually
dependent processes constitute one of the main sources for
operational risk. Indeed, an attempt to explore the consequences of
interactions for quantifying the capital buffer necessary to cover
operational risk \cite{KuehnNeu03} has provided major ingredients
for the approach to credit risk modelling started by Neu and one of
us \cite{NeuKuehn04}.

In the present paper we provide an analytic solution to the
dynamical description of counter-party risk within a heterogeneous,
functionally defined network of interacting firms (to be referred to
as economy in what follows), see e.g. \cite{Eglo03}, in the spirit
of \cite{NeuKuehn04}. We generalize the analysis of that study to
capture effects of cooperative as well as competitive business
relations within the economy, and we solve the model for a wide
degree of dilution of the network of economic dependencies in the
sense that we assume  each company in the net to have business
relations only with a (randomly chosen) subset of the full set of
companies. In the present investigation we will always consider the
case where the number of interaction partners of each company is
large, and for simplicity we shall restrict ourselves here to the
case where the graph defining economic connectivity is a Poisson
degree distributed Erd\"os-R\'enyi random graph \cite{Bollobas01}.
More realistic connectivity distributions reflecting the different
connectivity patterns of large and small players in an economy,
taking into account small-world effects and fat tailed degree
distributions \cite{Barabasi99} can be handled by methods similar to
those used in the present investigation \cite{Per+04, Hatchett05},
but will be studied in a separate paper.

As in \cite{NeuKuehn04}, the model parameters are unconditional and
conditional default probabilities, which may be thought of as being
obtained via a suitable rating procedure. One of the virtues of the
present analytic investigation is to highlight the fact that the
collective behaviour of the system, which ultimately determines the
the loss-distribution on an economy-wide scale, is fairly
insensitive to detail. That is, it does not depend on getting
individual dependencies correct, but only on the overall
distribution of of the unconditional and conditional default
probabilities.  Our main result is to demonstrate that the effects
of economic interactions --- while  relatively weak in typical
scenarios --- are pronounced in situations of economic stress, and
thus lead to a substantial fattening of the tails of loss
distributions even in large loan portfolios.

The remainder of the paper is organized as follows. In Sec.
\ref{sec:defs} we define our model, and specify the stochastic
setting for our analytic investigation. The relation between the
model parameters and conditional and unconditional default
probabilities as used in \cite{NeuKuehn04} is briefly reviewed to
make the paper self-contained. Sec. \ref{sec:heuristic} describes a
heuristic solution for the dynamical evolution of the fraction of
defaulted companies over a risk horizon of one year, starting from
an overall healthy situation, a scenario appropriate for the
analysis of credit risk. A formal solution in terms of a generating
function approach (GFA) \cite{Dominicis78}, which provides a full
justification for the heuristic solution is relegated to
\ref{app:gfa}. Both solution methods are based upon techniques
developed in the statistical mechanical analysis of dilute neural
networks \cite{Derrida87}. In Sec. \ref{sec:results} a phase diagram
distinguishing regions in parameter space in which economic
interactions can lead to a collective acceleration of the
economy-wide default rate in situations of economic stress from
regions where such acceleration is impossible is computed.
Distributions of annual fractions of defaulted companies as well as
loss distributions, both economy-wide and for finite loan-portfolios
are also computed and compared with simulations. Sec.
\ref{sec:conclusion} finally summarizes our findings and discusses
their main implications for the analysis of credit risk

\section{Model Definitions}
\label{sec:defs}

In this section we define a statistical model that attempts to capture
the effects of counter-party risk on credit contagion. In contrast to
approaches based on microeconomics, and in keeping with the framework
discussed in the introduction, we allow the firms' wealth,
macro-economic factors and interactions between firms to all be
described probabilistically. This is due to our focus on the
characteristic change in behaviour caused by examining interactions
between firms in the wider economy.

We analyse an economy which consists of $N$ firms. The state of each
firm $i$ at a given time $t$ is described by its `wealth' $W_{it}$
which we take to be the difference between its assets and its
liabilities. Accordingly, a company defaults, if its wealth $W_{it}$
falls below zero. As indicated in the introduction, we are
interested in the influence of economic partners on firms' own
performance, specifically on their default probability. We
understand the notions of firm and economic partner in a very wide
sense: a firm could be any economic entity, a manufacturer, a
service provider, a trader of goods or services, etc. It could also
be an individual (we shall often use the generic term `node' to
designate these entities). Two firms are  partners if the state of
one has a material effect upon the other, e.g. one is  the supplier
of the other, performs outsourced services, there exists substantial
loans or other financial commitments between the two, or they
compete in the  same market. For instance, if a major manufacturer
of PCs were to go out of business  tomorrow, this would inevitably
have a material impact on the economic performance  (i.e. wealth) of
other companies operating in the same industry sector. It would  on
the one hand lead to a deterioration in the financial viability of
most of the suppliers or service providers of the PC manufacturer in
question
--- including in particular its own work force! --- but on the other hand it
would improve the situation for competing producers of PCs, in that
they could profit by taking over a share of the defaulted company's
market.

We are interested in quantifying the effect of these interactions from the
perspective of a lending bank which would be required to set aside a sufficient
amount of capital to cover losses incurred by defaults of its obligors. Another
perspective might be that of a central bank, which would base monetary policy
decisions in part on their impact on expected default rates at an economy wide
scale. The typical risk horizon in these contexts would be one year.

For simplicity, we assume that that within the risk horizon of one
year node $i$ experiences an interaction-induced material change of
its wealth only if one of its business partners, say $j$, defaults.
In order to formalize this in a dynamical description, we  introduce
binary indicator variable $n_{jt}$ which indicates whether node $j$
is solvent at time $t$ ($n_{jt} = 0$) or has defaulted ($n_{jt} =
1$).

The value of the $i$th node's wealth at time $t$, $W_{it}$, is thus taken to be
of the form
\begin{eqnarray}
W_{it} = \vartheta_{i} - \sum_{j = 1}^n J_{ij} n_{jt} - \eta_{it}
\end{eqnarray}
Here $J_{ij}$ denotes the change in $i$'s wealth which would be
induced by a default of node $j$. One would have $J_{ij} >0$ if $j$
is a cooperative partner of $i$, whereas $J_{ij} <0$ if $j$ is a
competitor, while $J_{ij}=0$ if there is no direct influence  of $j$
on $i$. By $\vartheta_{i}$ we designate $i$'s initial wealth at the
beginning of the year, and  $\eta_{it}$ are random fluctuations
caused by both, {\em external\/} macro-economic factors  (an
expanding/shrinking economy, an oil price spike, market sentiment
etc), and {\em firm specific\/} actions or events.

We take the initial state of the economy to be a set of solvent
firms, $n_{i0}=0$ for all $i$, (one could view this as a definition)
and say that a firm defaults at time $t$ if $W_{it} < 0$. We define
our dynamics such that if a firm goes bankrupt it does not recover
within a risk horizon of one year, so the bankrupt state is
absorbing. Thus the dynamics of the firms state is given by the
equation
\begin{equation}
n_{it+1} = n_{it} + (1-n_{it})\Theta\Bigg(\sum_j J_{ij} n_{jt}-\vartheta_i
+ \eta_{it}\Bigg)
\label{eq:micro}
\end{equation}
where $\Theta(\ldots)$ is the Heavyside function. The time step in this dynamical
rule will be taken to represent one month.

We choose the $\eta_{it}$  to be Gaussian distributed. Without loss
of generality they can --- by suitably rescaling the $\vartheta_i$
and the $J_{ij}$ --- be chosen to have unit variance. We follow
widespread practice \cite{CM, Kealhofer98} to account for common
fluctuating macro-economic factors by choosing the $\eta_{it}$ to be
correlated for different $i$. This could be achieved by taking
$\eta_{it}$ to be of the form \be \eta_{it} = \sigma_i \xi_{it} +
\sum_{k=1}^K \beta_{ik} Y_{kt} \ee with uncorrelated Gaussian unit
variance white noises $\xi_{it}$ and $\{Y_{kt}\}$, the former
describing firm specific wealth fluctuations, whereas the latter
could account for the relative effects of fluctuations common to
industry sectors, regions, or countries, with prefactors $\sigma_i$
and $\beta_{ik}$ describing the relative importance of these
fluctuations on $i$. In what follows we restrict ourselves to a {\em
minimal variant\/} of this set-up by finally choosing
\begin{eqnarray}
\eta_{it} = \sqrt{\rho}~\eta_{0t} + \sqrt{1 - \rho}~\xi_{it}\ .
\label{eq:onefactor}
\end{eqnarray}
We shall simplify matters further by assuming that the common
economic factor $\eta_{0t}$ is slow and take it to be constant
$\eta_{0t}=\eta_{0}$ within a risk horizon of one year. One-factor
models of this type also feature in the regulatory framework laid
out in the Basel II accord \cite{BISII}.

None of the simplifying assumptions are necessary for our analysis
to go through; the generating function formalism given in
\ref{app:gfa} in particular can easily handle more general cases.
However, the simplified setting is sufficient to highlight the
important effects of interactions on credit risk, and it does lead
to a greatly simplified macroscopic description of the system, as we
will see in Sec. \ref{sec:heuristic}.

As for the $J_{ij}$ which describe the loss or gain of node $i$ due
to a default of node $j$, in the present paper we will investigate
them in a probabilistic setting. It will be useful to disentangle
the presence or absence of an interaction from its strength by
writing \be J_{ij} = c_{ij} \tilde J_{ij} \ee where $c_{ij} \in
\{0,1\}$ describes the absence or presence of a connection $j\to i$,
while $\tilde J_{ij}$ describes its magnitude, both of which are
assumed to be fixed. It is reasonable to assume that connectivity is
a symmetric relation, $c_{ij} = c_{ji}$, whereas there is no reason
to suppose symmetry of the magnitudes of mutual influences.
Considering the case of  a small supplier with one large company
taking the majority of its orders, if the larger company defaults
then the small supplier may well go bust too. However, if the small
supplier defaults then the large company is less likely to suffer
terminal financial distress, so in general $\tilde J_{ij} \ne \tilde
J_{ji}$.

Specifically, we assume a random connectivity pattern described by
\be P(c_{ij}) = \frac{c}{N} \delta_{c_{ij},1} +
\left(1-\frac{c}{N}\right) \delta_{c_{ij},0}\ \ , \ i < j\ \ , \
c_{ij}=c_{ji} \label{eq:PJij1} \ee and we will be interested in the
limit of a large economy (the thermodynamic limit), in which the
average connectivity $c$ of each node is itself large, $N \to
\infty$, $c \to\infty$. We will initially be concerned with the
extremely diluted regime, where  $c/N \to 0$, taking e.g. $c =
\mathcal{O}(\log(N))$. These assumptions have important consequences
for the structure of the graph defining the connectivities, namely
that each node feels the effects of a large number of other nodes
(so that limit theorems will allow us to describe the overall
effects of interactions) and, for the extremely diluted regime, that
there are only a finite number of loops of finite length even in the
infinite economy limit. The graph of interactions between companies
for finite $N$ is just an Erd\"os-R\'enyi random graph
\cite{Bollobas01}.

The magnitudes $\tilde J_{ij}$ of the interactions will also be
taken as fixed random quantities. In order to allow the taking of
the thermodynamic limit as described, the mean and fluctuations of
the $\tilde J_{ij}$ must scale in a suitable way with the
connectivity $c$. Quite generally, we must have \be \tilde J_{ij} =
\frac{J_0}{c} + \frac{J}{\sqrt{c}} x_{ij} \label{eq:PJij2} \ee in
which the $x_{ij}$ are zero-mean unit-variance random variables. The
scaling of mean and variance of the $\tilde J_{ij}$ is given by the
parameters $J_0$ and $J$ respectively. If $J_0>0$ there will be a
net cooperative tendency within the economy, which seems to be a
reasonable assumption. Finally, we will need to assume that all
moments of the $x_{ij}$ are finite and we will choose the  $x_{ij}$
 to be {\em independent in pairs\/}
\begin{eqnarray}
\overline x_{ij}=0\ , \qquad \overline {x^2_{ij}}=1\  ,  \qquad
\overline{x_{ij} x_{ji}}=\alpha\ , \qquad \overline{x_{ij}x_{kl}} =0\ \mbox{otherwise}\ .
\label{eq:PJij3}
\end{eqnarray}
The parameter $\alpha$, ($-1\le \alpha\le 1$) describes the degree of correlations
between $J_{ij}$ and $J_{ji}$. Strictly symmetric interactions are obtained only for
$\alpha=1$

At this point let us briefly recall that, after rescaling as
described, the model parameters $\vartheta_i$ and $J_{ij}$ have a
clear meaning in terms of unconditional and conditional default
probabilities \cite{NeuKuehn04}. We denote by ${\bm n}_t$ the values
of all indicator variables in the economy at time $t$, assuming
$n_{it}=0$. Then by integrating over the unit variance Gaussian
$\eta_{it}$ in (\ref{eq:micro}) one obtains the conditional
probability for node $i$ to default within a month given a
configuration ${\bm n}_t$  of non-defaulted and defaulted firms in
the economy at time $t$ as,
$$
{\rm Prob}\left.\Big(n_{it+1}=1 \right|{\bm n}_t\Big)
= \Phi\Big(\sum_{j} J_{ij}\, n_{jt} - \vartheta_i \Big)
$$
with $\Phi(x)= \frac{1}{2}[1+\erf(x/\sqrt 2)]$ the cumulative normal
distribution. Thus the unconditional probability $p_{i}$ of default
of $i$ within a month in an otherwise healthy economy and the
conditional probability $p_{i|j}$ for a default of $i$ within a
month, given $j$ and only $j$ has defaulted before are given by \bea
p_{i}   &=& \Prob\left.\Big(n_{it+1}=1 \right|\{n_{it} =  0\}\Big) =
\Phi(-\vartheta_i)\ ,
\label{pd_uncond}\\
p_{i|j} &= &\Prob\left.\Big(n_{it+1}=1
\right|n_{jt}=1, \{n_{k(\ne j)t}=0\}\Big)
 =  \Phi\Big(J_{ij}-\vartheta_i\Big) \ .
\label{pd_cond} \eea These relations may be inverted to express the
model parameters in terms of conditional and unconditional default
probabilities --- quantities that would be estimated in a rating
procedure --- as \be \vartheta_i = - \Phi^{-1}(p_{i})\quad ,\qquad
J_{ij} = \Phi^{-1}(p_{i|j}) - \Phi^{-1}(p_{i}) \ . \ee

While characterising the default of companies is of interest, our
primary concern is to examine the distribution of losses accrued
over our one year time frame, both in the economy at large, and in a
portfolio made up of a finite number of firms within the economy. We
assume that the losses caused by default are independent of the
month of default, and then examine two different cases. The first
simpler case is that the losses at firm $i$, given that firm $i$
defaults, are uncorrelated with any other variables. The second,
perhaps more interesting case, is that the losses at firm $i$ are
random but are correlated with the initial monetary reserves
$\vartheta_{i}$. The intuitive reasons are that if a firm has more
cash, then the default is less anticipated, and thus will be less
priced in by the market; or the firm has larger credit lines and so
will default on a larger amount; and finally the firm is likely to
be larger, and hence cause a larger loss.

\section{Heuristic Solution}
\label{sec:heuristic} In the present section we show that our model
has a relatively simple solution that can be obtained by qualitative
probabilistic reasoning, appealing to statistical limit theorems.
This solution turns out to be exact, as we show using a more
involved generating function formalism in \ref{app:gfa}. Both types
of argument have been developed in the analysis of the statistical
mechanics of disordered systems, and in particular neural network
models \cite{Derrida87}, while for a more general introduction to
emergent collective behaviour see e.g.
\cite{Phasetrans}. Recall the microscopic dynamics as defined by
(\ref{eq:micro}). The complications are due to the interactions
between firms, namely that the state of a given firm $i$ at time
$t$, depends on the state of the neighbours of $i$ for times
$t^\prime < t$ which in turn depend on $i$ at times $t^{\prime
\prime} < t^\prime < t$. In general this feedback prohibits
straightforward analysis, and indeed, it led Jarrow and Yu
\cite{Jarrow01} to eliminate this feedback explicitly by considering
an economy of two types of firms: primary firms whose default
depended only on macro-economic factors and secondary firms whose
default depended on macro-economic factors and the default of primary
firms. However, due to the specific structure of our model we are
able to push the analysis further. By definition, the overall effect
of interaction terms on company $i$ at time $t$ is given by the
local field $h_{it}=\sum_j J_{ij} n_{jt}$. From the statistics of
the interactions $J_{ij}$ given by (\ref{eq:PJij1})-(\ref{eq:PJij3})
we see that each firm $i$ is connected to, on average, $c$ other
firms. Since we consider the large $c$ limit, this means that we
could evaluate the statistics of $h_{it}$ by appeal to the law of
large numbers and the central limit theorem if the contributions to
$h_{it}$ were independent, or at least sufficiently weakly
correlated.

At first sight we cannot expect this condition to hold if we have
some degree of symmetry in the interactions, i.e. for $\alpha \ne
0$, even in the extremely diluted regime. Note that
there are two ways in which the $n_{jt}$ of the neighbours
interacting with $i$ may become correlated through the dynamics:
either they influence each other through firm $i$, or not through
firm $i$ but though some loop of interactions in the economy. In the
extremely diluted regime, correlations between the neighbours $j$ of
$i$ cannot build up in finite time (within the risk horizon) via
loops not involving $i$, since due to the scaling almost all loops
are very long. With symmetry in the interactions, correlations
between the $n_{jt}$ could in principle be induced by the dynamics
of $i$. However, as long as $n_{it} = 0$, the $n_{jt}$ clearly
cannot influence each other through site $i$, whereas once $n_{it} =
1$, then firm $i$ is in the absorbing state, and correlations it
induces on the dynamics of its neighbours, have become irrelevant
for its own microscopic dynamics (\ref{eq:micro}). Thus limit
theorems can be used after all to solve the macroscopic dynamics of
the system, despite a possible symmetry in the interactions.

Returning to the dynamical evolution equation (\ref{eq:micro}), we observe
that the coupling of a node to the economy is via the local field
\be
h_{it} = \sum_j J_{ij} n_{jt}= \frac{J_0}{c}\sum_j c_{ij} n_{jt}
 + \frac{J}{\sqrt{c}} \sum_j c_{ij} x_{ij} n_{jt}\ ,
\label{eq:locf}
\ee
which is a sum of random quantities (with randomness both due to the Gaussian
fluctuating forces (the $\{\eta_{it}\}$, respectively the $\{\xi_{it}\}$), and
due to the heterogeneity of the environment). The first contribution is a
sum of terms of non-vanishing average. By the law of large numbers this sum
converges to the sum of averages in the large $c$ limit,
$$
h^0_{it}\equiv \frac{J_0}{c} \sum_j c_{ij} n_{jt} \rightarrow \frac{J_0}{c}
\sum_j \overline {c_{ij} \bra n_{jt}\ket } \simeq \frac{J_0}{c} \sum_j \overline {c_{ij}}~\overline {\bra n_{jt}\ket} =  J_0\frac{1}{N} \sum_j \overline {\bra n_{jt}\ket}
$$
in which angled brackets $\bra \dots \ket$ denote an average over
the fluctuating forces, and the overbar $\overline{(\dots)}$ an
average over the $J_{ij}$, i.e., the  $c_{ij}$ and the $x_{ij}$. An
approximation is made by assuming negligible correlations between
the $c_{ij}$ and the $\bra n_{jt}\ket$ induced by the heterogeneity
of the interactions. The second contribution to (\ref{eq:locf}) is a
sum of random variables with zero mean, which we have argued are
sufficiently weakly correlated for the central limit theorem to
apply for describing the statistics of their sum. Thus the sum
$$
\delta h_{it} \equiv \frac{J}{\sqrt c} \sum_j c_{ij} x_{ij} n_{jt}
$$
is a zero-mean Gaussian whose variance follows from
\bea
\overline{\bra (\delta h_{it})^2\ket}&=& \frac{J^2}{c} \sum_{jk}
\overline{c_{ij}c_{ik} x_{ij} x_{ik} \bra n_{jt}n_{kt}\ket} \simeq
\frac{J^2}{c} \sum_{jk} \overline{c_{ij}c_{ik} x_{ij} x_{ik}}
\overline{\bra n_{jt}n_{kt}\ket} \nonumber\\
&=& J^2 \frac{1}{N} \sum_j \overline{\bra n_{jt}\ket}\nonumber \eea
An approximation based on assuming negligible correlations has been
made as for the first contributions. Thus the local field $h_{it}$
is a Gaussian with mean $h^0_{it}$ and variance $\overline{\bra
(\delta h_{it})^2\ket}$ both scaling with the average fraction of
defaulted nodes in the economy. By the law of large numbers this
average fraction will be typically realized in a large economy, i.e.
we have \be m_t = \frac{1}{N} \sum_j n_{jt} \rightarrow \frac{1}{N}
\sum_j \overline{\bra n_{jt}\ket} \ee in the large $N$ limit. The
dynamics of the fraction of defaulted nodes then follows from
(\ref{eq:micro}), \be \hspace{-2cm} m_{t+1}= \frac{1}{N} \sum_i
n_{it+1} = m_t + \frac{1}{N} \sum_i \big (1-n_{it}\big)
\Theta\Big(h_{it}-\vartheta_i  +\sqrt\rho \eta_{0} +\sqrt{1-\rho}~
\xi_{it} \Big)\ , \label{eq:macro1} \ee where the one factor noise
model (\ref{eq:onefactor}) has been used.

The sum in (\ref{eq:macro1}) is evaluated as a sum of averages over
joint $n_{it}$, $h_{it}$, and $\xi_{it}$ distribution by the law of
large numbers. We exploit the fact that $n_{it}$, $\xi_{it}$ and
$h_{it}$ are uncorrelated. Noting that the sum
$h_{it}+\sqrt{1-\rho}~ \xi_{it}$ is Gaussian with mean $J_0 m_t$ and
variance $1-\rho + J^2 m_t$, and taking into account that
$n_{it}$-averages, depend on $i$ through $\vartheta_i$, $\bra
n_{it}\ket=\bra n_{t}\ket_{(\vartheta_i)}$, we find
$$
m_{t+1}= m_t +\frac{1}{N} \sum_i \frac{1- \bra n_{t} \ket_{(\vartheta_i)}}{2}\left[1+
\erf\left(\frac{J_0 m_t + \sqrt\rho ~\eta_{0}-\vartheta_i}
{\sqrt{2(1-\rho + J^2 m_t)}}\right)\right]
$$
This version can be understood as an average over the $\vartheta$
distribution
$$
p(\vartheta) = \frac{1}{N} \sum_i \delta(\vartheta - \vartheta_{i})\ ,
$$
which maps onto a distribution of unconditional default
probabilities as discussed above. Denoting that average by $\bra
\dots \ket_\vartheta$ we finally get the following evolution
equation for the macroscopic fraction of defaulted companies in the
economy
 \be m_{t+1} = m_t + \left\bra \frac{1- \bra n_t
\ket_{(\vartheta)}}{2}\left[1+ \erf\left(\frac{J_0 m_t + \sqrt\rho
~\eta_{0}-\vartheta} {\sqrt{2(1-\rho + J^2
m_t)}}\right)\right]\right\ket_{\vartheta} \label{eq:macro2} \ee

We have thus an explicit dynamic equation for the macroscopic fraction of
defaulted nodes in the economy. It involves first propagating $\vartheta$-dependent
default probabilities via
\be
\bra n_{t+1} \ket_{(\vartheta)} = \bra n_{t} \ket_{(\vartheta)}
+ \frac{1- \bra n_t \ket_{(\vartheta)}}{2} \left[1+
\erf\left(\frac{J_0 m_t + \sqrt\rho ~\eta_{0}-\vartheta}
{\sqrt{2(1-\rho + J^2 m_t)}}\right)\right]\ ,
\label{eq:macro3}
\ee
which depends only on $m_t$, thereafter performing an integral over the $\vartheta$ distribution to obtain the updated fraction $m_{t+1}$ of defaulted nodes given in
(\ref{eq:macro2}).

The heuristic solution of the macroscopic dynamics
(\ref{eq:macro2}), (\ref{eq:macro3}) presented here is based on
independence assumptions which are not easily justified in a
rigorous way via the probabilistic reasoning presented above.
However, the solution is supported in full detail by an exact
analysis based on generating functions presented in \ref{app:gfa}.

\section{Results}
\label{sec:results}

In the present section we explore the consequences of our theory. We
studied the dynamics and computed loss distributions for an economy
in which the parameters $\vartheta_i$ determining unconditional
monthly default probabilities according to (\ref{pd_uncond}) are
normally distributed with mean $\vartheta_0=3$, and variance
$\sigma^2_\vartheta=0.01$ so that typical monthly default
probabilities are in the 5 $\times 10^{-4}$ range. Except when
stated otherwise we shall use $\rho=0.15$ for the parameter
describing the relative importance of economy-wide fluctuations, a
value that is considered to be in an economically acceptable range.

In Fig 1a we show the evolution of the {\em typical\/} fraction of defaulted firms over a
risk horizon of 12 months for various settings of the interaction parameters $J_0$ and $J$;
the typical fraction is computed by choosing the most-probable value $\eta_0=0$ for
the economy-wide influence on the dynamics. Fig 1b shows the probability density of the
end of year fraction of defaulted firms driven by fluctuations in economic conditions.
It is obvious that interactions cause a significant fattening of the tail of the  density
at large values of this fraction, which is a clear indication of the significance of
counter-party risk in particular in situations of economic stress.

\begin{figure}[h]
\begin{center}
\epsfig{file=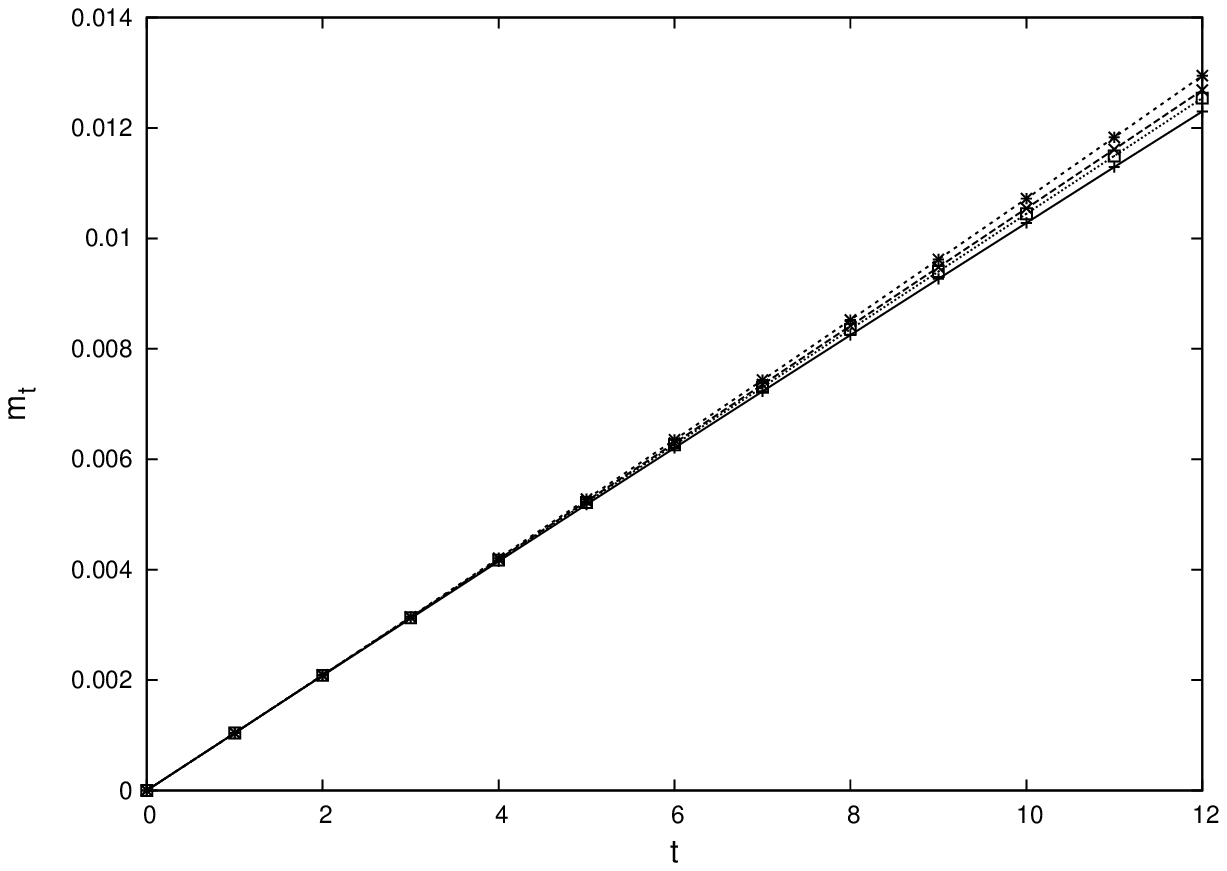,width=7.5cm}
\hfill
\epsfig{file=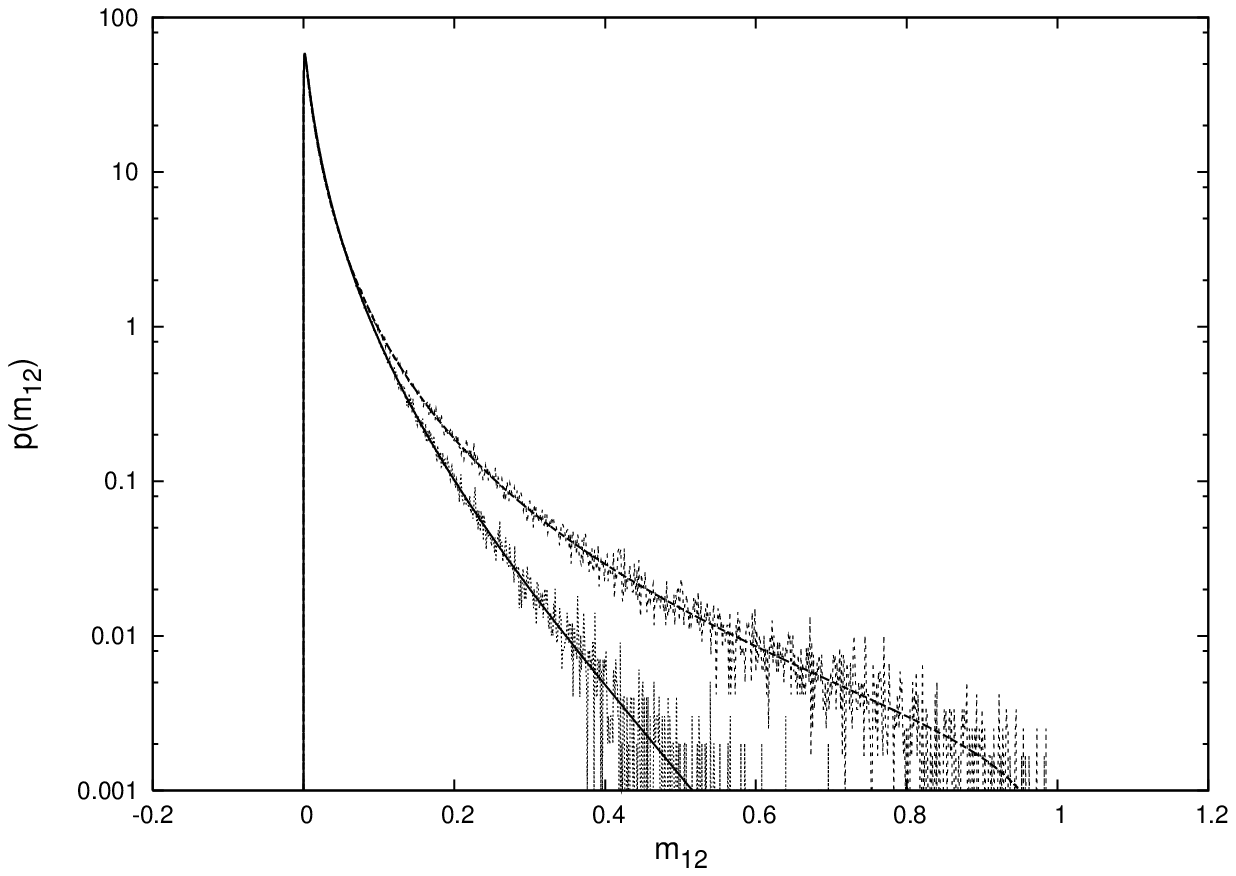,width=7.5cm}
\end{center}
\caption[]{Left: Typical fraction of defaulted companies as a function of
time for $(J_0,J)= (0,0),(1,0),(0,1)$, and (1,1) (bottom to top). Right: Distribution
of the fraction of defaulted companies at $t=12$ for $(J_0,J)= (0,0)$, and (1,1).
Analytic curves are overlaid with results of a simulation in which the distribution is
obtained  by computing the fraction of defaulted companies for randomly sampled $\eta_0$.}
\end{figure}

In order to assess whether interactions can lead to a collective
acceleration of the rate of defaults we look at the discrete second
derivatives
$$
\Delta_t = m_{t+1}+m_{t-1}- 2m_t
$$
which are always negative for the non-interacting system and maximal
at $t=1$, irrespectively of $\eta_0$. Interactions can lead to a
collective acceleration of the rate of defaults signified by the
possibility that the $\Delta_t$ may become positive in unfavourable
economic conditions. We define the region in parameter space in
which collective acceleration of default rates can occur by the
condition that \be {\rm Prob}\Big\{\Delta_t > 0 \Big\} > 0 \ee for
some $t$, with $1 <t< 11$. The concavity of the error function for
positive arguments entails that effects of collective acceleration
are always strongest at $t=1$. Evaluating this condition for various
values of the parameter $\rho$ describing the coupling to the
overall economy, we get lines shown in the phase diagram Fig 2a.
Note that the influence of $\rho$ is very weak in interesting region
of low $\rho$ values. Note also that $\Delta_1$ values are {\em
typically} positive but very small in the region near the phase
boundaries shown, as illustrated in In Fig 2b where we exhibit the
distribution of discrete second derivatives at $t=1$, just inside
the phase where acceleration of default rates is observed. The tail
of negative $\Delta_1$ is found to extend to significantly larger
values.

\begin{figure}[h]
\begin{center}
\epsfig{file=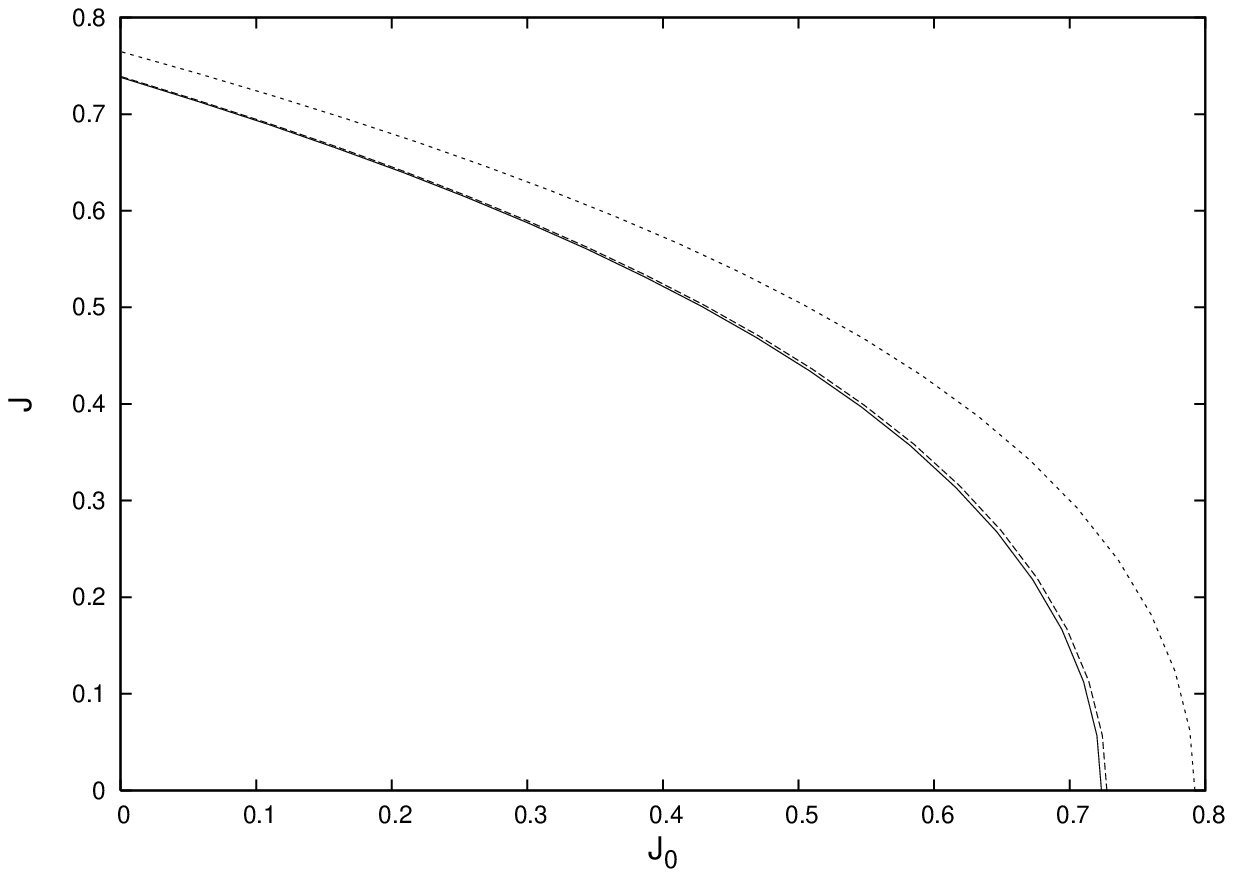,width=7.5cm}
\hfill
\epsfig{file=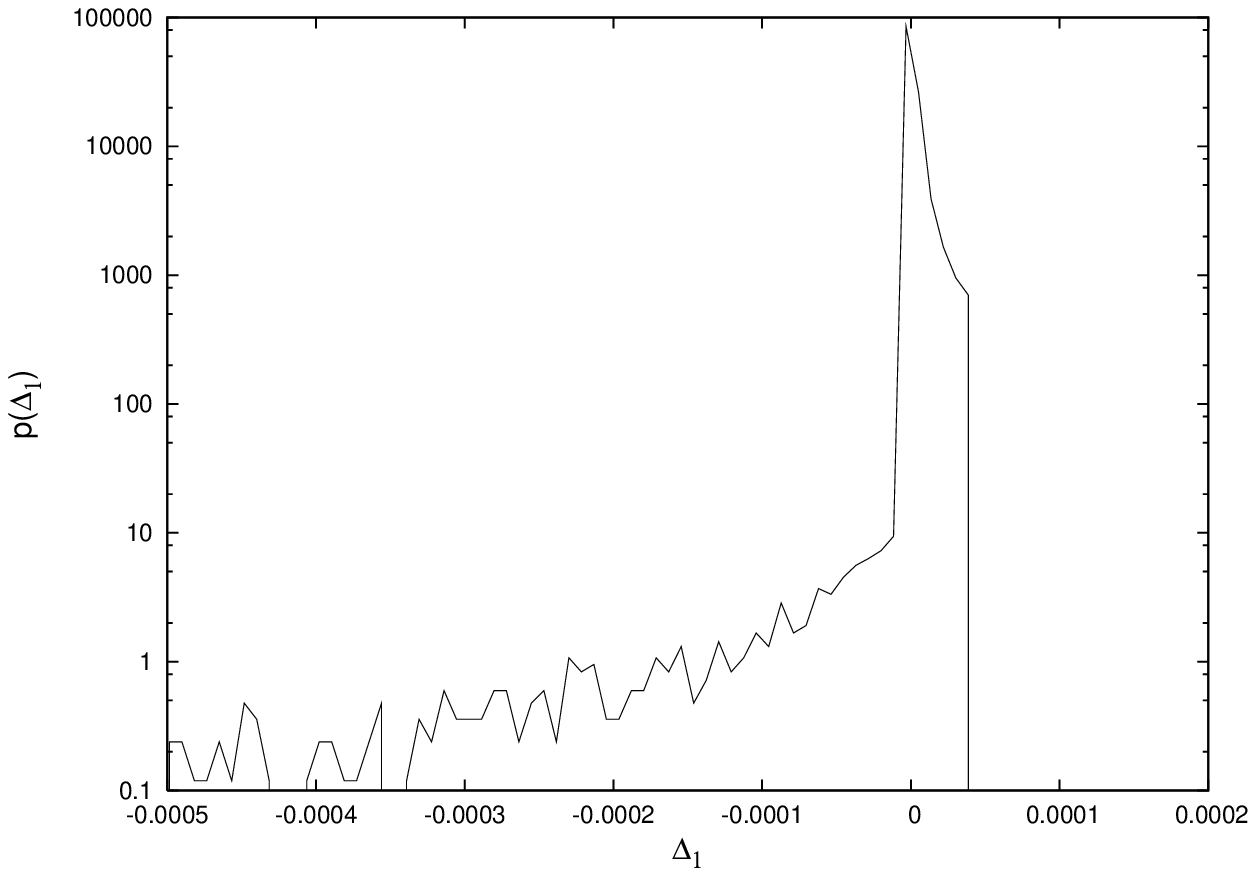,width=7.5cm}
\end{center}
\caption[]{Left: Phase boundaries separating regions without collective
acceleration of default rates from regions where acceleration occurs, for
$\rho=0.15$, 0.3 and 0.8 (bottom to top). Distribution of discrete second
derivative $\Delta_1=m_2+m_0- 2m_1$ just within the phase with accelerating
default rates for $(J_0,J)=(0.5,0.5)$ (right).}
\end{figure}

The quantity of central importance from the point of view of credit
risk analysis is of course the distribution of losses. Let $\ell_i$
denote the loss that would be incurred by a default of node $i$.
Then the loss per node for a given state $\eta_0$ of the economy is
\be L(\eta_0) = \frac{1}{N} \sum_i n_{i12} \ell_i \label{Leta0} \ee
where $\ell_i$ is randomly sampled from the loss distribution for
node $i$. We assume that the $\ell_i$ are independent of the
stochastic evolution. In the large system limit, the loss per node
at given value of $\eta_0$ describing the influence of the overall
economy is a non-fluctuating quantity, as it is itself an
(empirical) average taken over an (infinitely) large system, 
\be
L(\eta_0)= \lim_{N\to\infty} \frac{1}{N}\sum_i n_{12}(\vartheta_i)
\ell(\vartheta_i) = \int d\vartheta p(\vartheta) \langle
n_{12}\rangle_{(\vartheta)} \overline \ell(\vartheta) 
\ee 
by the law of large numbers, where $\overline \ell=\overline \ell(\vartheta)$
is the mean of the loss distribution for a node. If loss
distributions were identical for each node, with means independent
of default probabilities $\overline \ell(\vartheta)=\ell_0$, then
the distribution of losses driven by the fluctuations of the
economic stresses would simply replicate the distribution of the
fraction of defaulted firms.

The situation is different if loss-distributions are correlated with
default probabilities. As an example we consider the case where
average losses are inversely proportional to the unconditional
default probabilities $p_d(\vartheta_i)=p_i$ introduced in
(\ref{pd_uncond}). \be \overline \ell(\vartheta)
=\frac{\ell_0}{\varepsilon + p_d(\vartheta)} \label{scalelbar} \ee
with a parameter $\varepsilon$ introduced as a regularizer to
prevent divergence as $p_d\to 0$. That is, the contribution to the
total losses incurred by defaulting firms with different
unconditional default probabilities is approximately uniform over
the default probabilities. In our model we have \be p_d(\vartheta) =
\frac{1}{2}[1-{\rm erf}(\vartheta/\sqrt 2)] \ee Fig. 3 shows loss
distributions for such a situation. The analytic curves are computed
by noting that the losses per node are monotone increasing functions
of $\eta_0$ which is itself $\cN(0,1)$. Integrated loss
distributions are thus simply obtained using error functions
$$
\Prob\big[L(\eta_0) \le L\big] =  \frac{1}{2}[1+{\rm erf}(\eta_0(L)/\sqrt 2)]
$$
where $\eta_0(L)$ is the $\eta_0$-value giving rise to loss $L$ per
node. The probability density function is obtained via a single
numerical differentiation. 

\begin{figure}[h]
\begin{center}
\epsfig{file=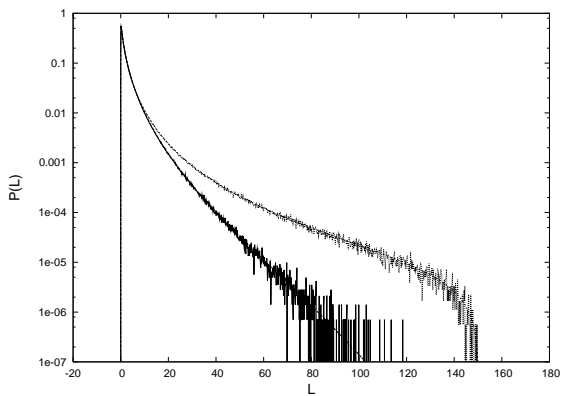,width=7.5cm}
\end{center}
\caption[]{Loss-distribution per node for the infinite system with
$\overline \ell(\vartheta) = 1/(\varepsilon + p_d(\vartheta))$ at
$\varepsilon=0.005$. Analytic curves are overlaid with simulation
results. Lower curve: non-interacting system, upper curve: interacting 
system with $(J_0,J)=(1,1)$.}
\end{figure}

It would be of some interest to know whether finite sample
fluctuations could possibly upset the picture seen so far. To study
this issue we look at the losses per node of a finite sample
randomly drawn from the nodes of a large economy, \be L_M(\eta_0)=
\frac{1}{M}\sum_{i=1 }^M n_{12}(\vartheta_i) \ell(\vartheta_i) \ee
Writing this as \bea L_M(\eta_0) &=& \frac{1}{M}\sum_{i=1}^M
\Big\langle \langle n_{12}\rangle_{(\vartheta)}
\overline\ell(\vartheta)\Big\rangle_\vartheta \nonumber\\
& & + \frac{1}{M}\sum_{i=1}^M \Big(
 \langle n_{12}\rangle_{(\vartheta_i)} \overline\ell(\vartheta_i)
 -\Big\langle \langle n_{12}\rangle_{(\vartheta)} \overline\ell(\vartheta)
\Big \rangle_\vartheta
\Big )\nonumber\\
& & + \frac{1}{M}\sum_{i=1}^M \Big(n_{12}(\vartheta_i) \ell(\vartheta_i)
- \langle n_{12}\rangle_{(\vartheta_i)} \overline\ell(\vartheta_i)
\Big ) \nonumber\\
&=& L_(\eta_0)+
L_2(\eta_0,\{\vartheta_i\})+L_3(\eta_0,\{\vartheta_i\},\{\phi_{it}\})
\eea we see that it has three components. The first is simply the
expectation value describing the loss per node at given $\eta_0$ in
an infinite system, the third, $L_3$ has zero mean and is expected
to be Gaussian at large $M$, with variance scaling as $M^{-1}$
describing the noise induced fluctuations about the average for a
given collection of $\{\vartheta_i\}$, while the second, $L_2$ ---
also a a zero mean Gaussian of variance scaling as $M^{-1}$ at large
$M$ --- describes the finite sample fluctuations of this average.
Since the collection of $\{\vartheta_i\}$ is fixed, these Gaussians
are correlated for different $\eta_0$. While an analytic evaluation
of the loss distribution may still be feasible in principle, it
would become very involved in practice.

An approximation to the finite size computation is obtained by
assuming that losses per node at given $\eta_0$ are Normally
distributed about their infinite system $\eta_0$-dependent mean with
[combining $L_2$ and $L_3$] variance (also $\eta_0$-dependent) \be
\sigma_M^2=M^{-1}\Bigg(\Big\langle\langle
n_{12}\rangle_{(\vartheta)}
\overline{\ell^2(\vartheta)}\Big\rangle_\vartheta -
\Big\langle\langle n_{12}(\vartheta)\rangle \overline \ell
(\vartheta)\Big\rangle_\vartheta^2\Bigg)\ , \ee which is an annealed
approximation which ignores that the parameters of the individual
loss distributions of the nodes in question remain fixed. The
results of an evaluation along this line are shown in Fig 4, using
the scaling (\ref{scalelbar}) of average losses used above; the
approximation suggests that finite size fluctuations give rise to
fatter tails in the loss distributions, though the effect is
negligible in the interacting system except at the extreme end of
the loss distribution, and small but a bit more pronounced in the
non-interacting system. A comparison with a simulation shows that
the effects of fluctuations are slightly underestimated in our
approximation. Note, however, that loan portfolios of typical banks
usually contain orders of magnitude more debtors than the $M=100$
considered in the present example.
\begin{figure}[h]
\begin{center}
\epsfig{file=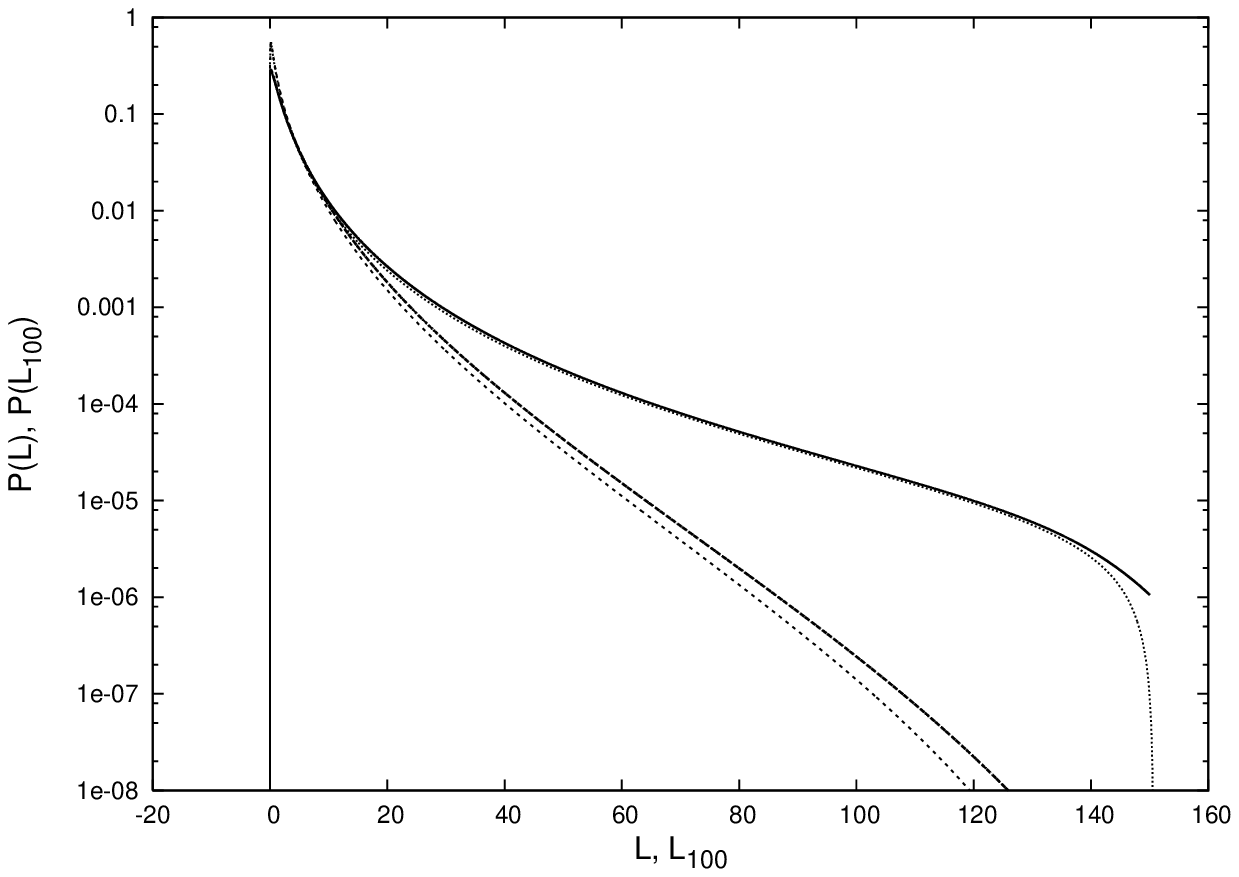,width=7.5cm}
\hfill
\epsfig{file=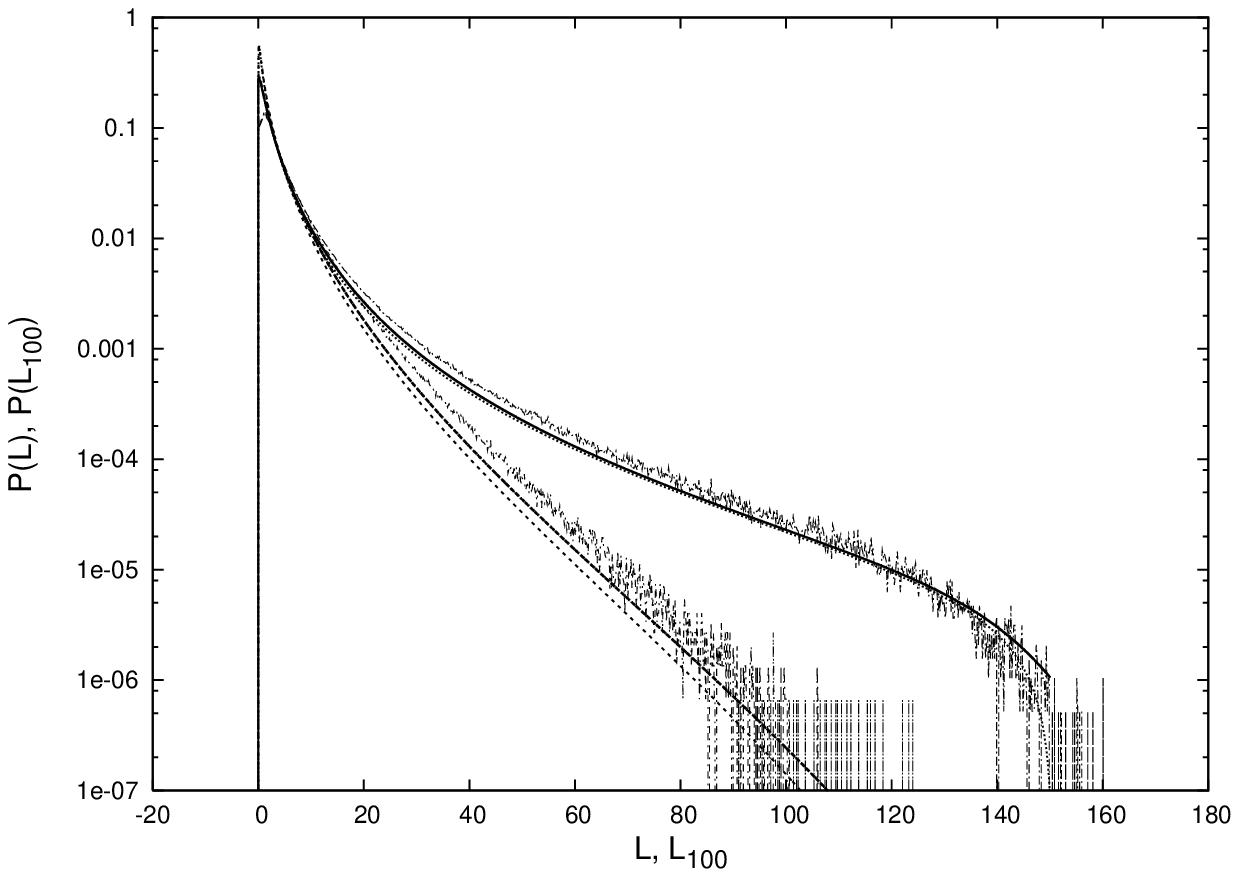,width=7.5cm}
\end{center}
\caption[]{Loss-distribution per node in the large system system limit with $\overline
\ell(\vartheta) = 1/(\varepsilon + p_d(\vartheta))$ at $\varepsilon=0.005$ and
for a finite sample of $M=100$ companies randomly taken from the ensemble. For
the finite sample, individual loss distributions are taken to be flat in the range $[0, 2\overline \ell(\vartheta]$. Right: The same analytic curves, overlaid with a
simulation result for the finite sample. Lower curves correspond in both cases to the non-interacting system, upper curves to the interacting system with $(J_0,J)=(1,1)$. }
\end{figure}

Let us finally look at the so-called {\em Value at Risk} in terms of
which the capital buffer that banks are required to hold to cover risk 
is often expressed. It is defined as 
\be 
{\rm VaR}_{q}=(Q_q[L]-\langle
L\rangle)~\rme^{-rT} \ee 
in which $Q_q[L]$ is the $q$-quantile of the loss distribution at time 
$T$, i.e. the loss that is not exceeded with probability $q$
$$
{\rm Prob}(L \le Q_q[L]) = q \ ,
$$
while $\bra L\ket$ is the average loss, and $r$ denotes a risk free
interest rate. To highlight the effects introduced by economic interactions 
we take the ratio of the value at risk computed for economies with and 
without functional economic interactions ${\rm VaR}/{\rm VaR}_0$, both
computed at confidence level $q=0.999$ as required by the Basel II regulations 
\cite{BISII}. Taking this ratio also eliminates the dependence on the interest 
rate $r$. 

In Fig 5 we display this ratio alongside with analogous ratios of average losses
$\bra L\ket/\bra L\ket_0$ for comparison.

\begin{figure}[h!]
\begin{center}
\epsfig{file=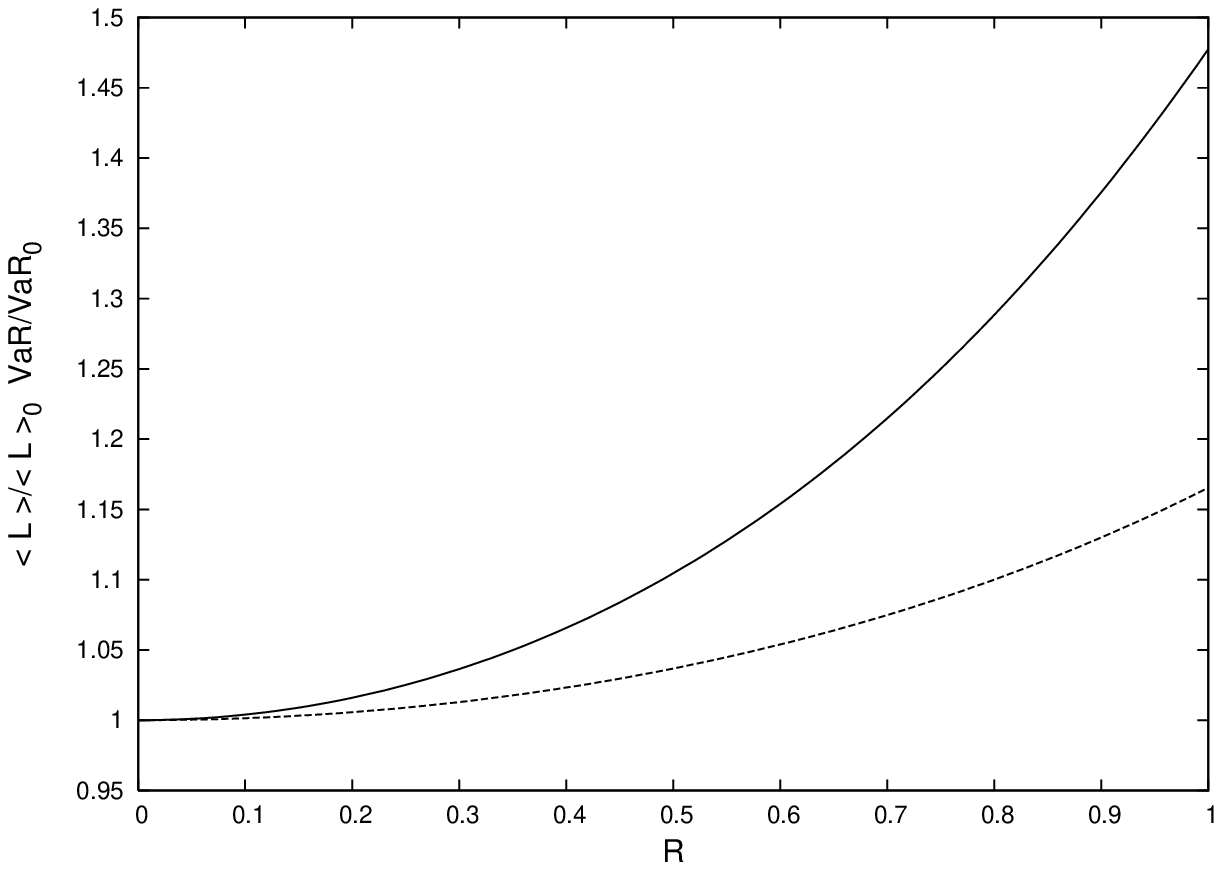,width=7.5cm}
\hfill
\epsfig{file=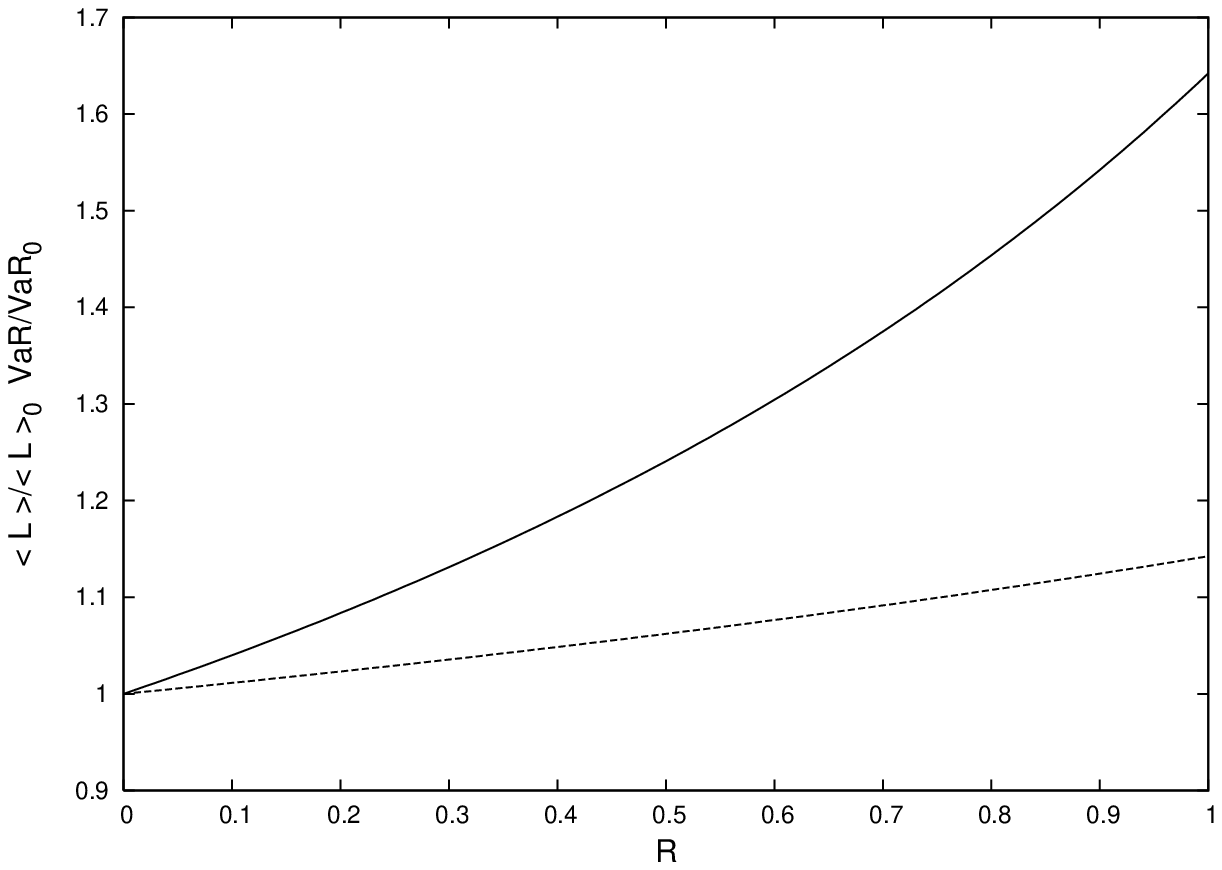,width=7.5cm}
\hfill
\epsfig{file=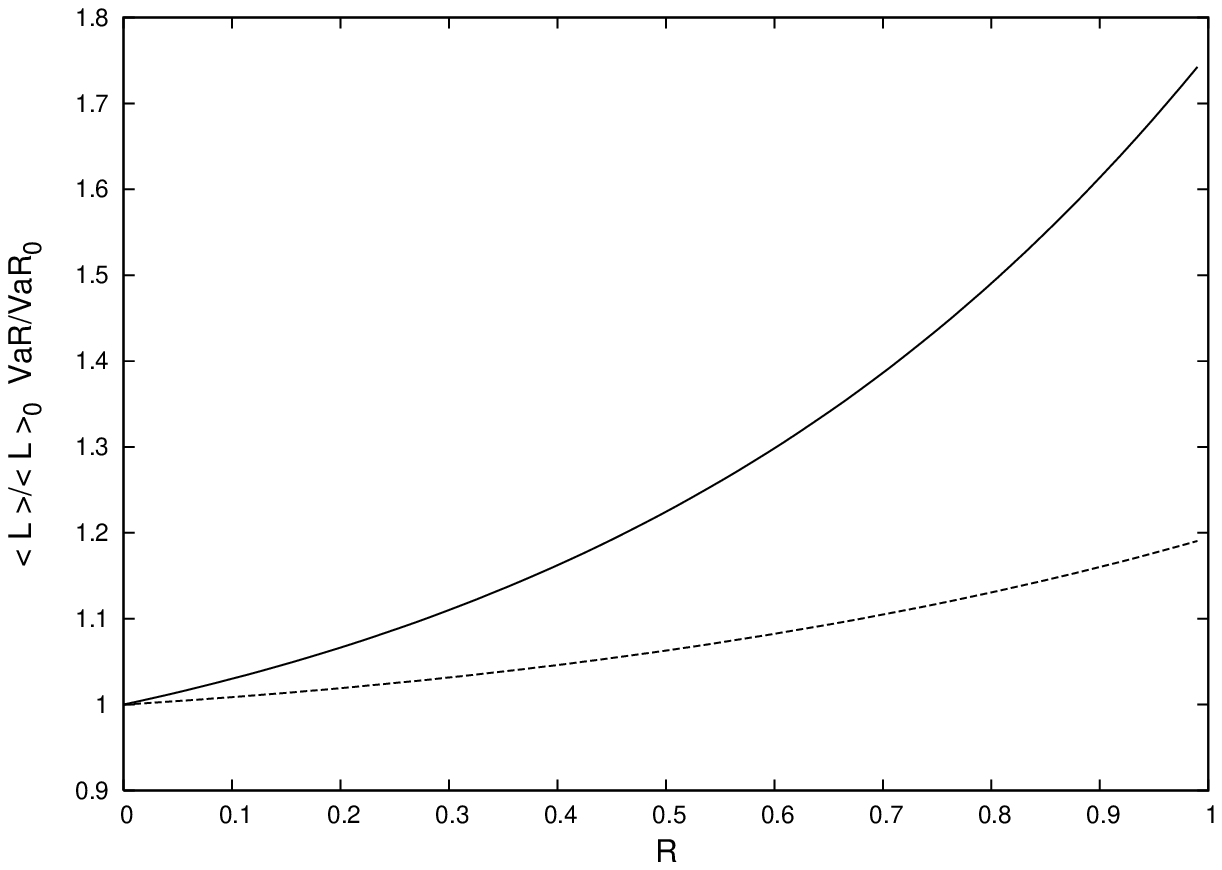,width=7.5cm}
\end{center}
\caption[]{Ratio of Value at Risk for systems with and without
functional interaction as a function of the strength of the
interaction (upper curves). The analogous ratio for average losses
is also shown in each case (lower curves). The curves shown in the
three figures are evaluated along straight lines in the $J_0-J$
plane, and the parameter $R$ measures a distance from the origin
$R=\sqrt{J_0^2+J^2}$. The three figure correspond to the lines
$J_0=0$ (upper left), $J=0$ (upper right) and $J_0/J=1$ (lower).}
\end{figure}

As perhaps may be anticipated in view of results displayed in Figs 3
and 4, the Value at Risk is significantly more sensitive to
functional interactions in an economy,  than the average losses are.
This is understandable, as VaR probes the tails of loss
distributions, while average losses will be determined mostly by
typical results.

For the results displayed in Fig 5, unconditional default probabilities
where {\em not\/} adjusted with the strength of the interactions so 
as to keep the average annual default probability constant. In Fig 
6, therefore, we take this extra step, displaying an analogous ratio of
the Value at Risk of interacting and non-interacting economies, where now
unconditional default probabilities in the interacting system are adjusted
in such a way that the average (interaction-renormalized) default probability
stays constant --- at the level chosen for the non-interacting system.
To keep matters simple, a homogeneous portfolio, with firm-independent 
unconditional default probabilities and firm independent average losses
was chosen. Clearly the interaction-induced enhancement of the Value at
Risk is rather close to the corresponding enhancement computed without 
adjustment of the unconditional default probabilities. 

\begin{figure}[h]
\begin{center}
\epsfig{file=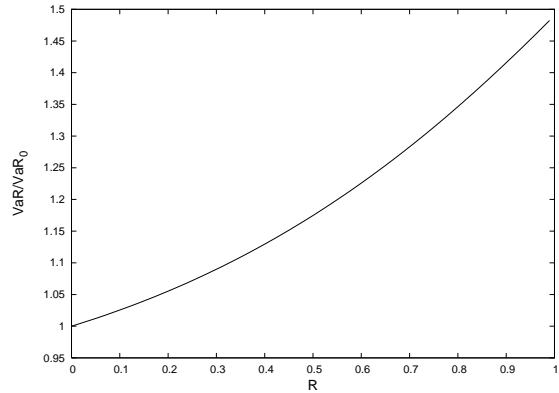,width=7.5cm}
\end{center}
\caption[]{Value at Risk relative to Value at Risk in systems
without functional interaction as a function of the strength 
$R=\sqrt{J_0^2+J^2}$ of the interaction for $J_0/J=1$, with unconditional
default probabilities in the interacting system adjusted as a function 
of the strength $R$ of the interaction to keep the annual average default
probability constant (in the present case at a value close to 3\%.). As in
Fig. 5, the VaR ratio is computed at $q=0.999$.}
\end{figure}

To summarize, the capital buffer that banks are required to hold according
to the Basel II regulations \cite{BISII} to cover credit risk is {\em significantly 
underestimated\/} when interaction effects in an economy are not taken into account. 
It is important to note that this is true already in the regime in which interactions
are too weak to cause an overall acceleration of default rates, as can be 
seen by comparing the phase diagram Fig. 2a with results for the Value at
Risk displayed in Figs. 5 and 6.

\section{Conclusion}
\label{sec:conclusion}

In conclusion, we have studied the effects of economic interactions
on credit risks. Though non-equilibrium initial conditions and the
fact that the credit-risk problem has an absorbing state would at
first sight appear to complicate the analysis, we found, quite to
our own surprise, that in particular the presence of the absorbing
state simplifies the analysis considerably, as it removes the
non-Markovian effects in the macroscopic dynamics that would
otherwise be present in systems with some degree of symmetry in the
interactions. While the limit of extreme dilution simplified the
reasoning within the heuristic solution, we saw in the generating
function analysis that the assumption of extreme dilution could be
dispensed with. So although the rather heavy machinery of
non-equilibrium disordered systems theory is required to rigorously
treat the model (due to asymmetry in the inter-firm dependencies and
the initial conditions), the resulting effective single-firm process
is remarkably simple. This has obvious practical benefits in terms
of computational efficiency.

We have seen that the effects of economic interactions are
relatively weak in typical economic scenarios, but they are
pronounced in situations of economic stress, and thus lead to a
substantial fattening of the tails of loss distributions in large
loan portfolios. This leads to significant increases in the Value 
at Risk, i.e. the capital that must be held as a loss buffer, when 
compared to the non-interacting theory. Importantly, this conclusion 
remains valid even in the case where there is no overall acceleration 
in default rates, c.f. Fig. 2a and Figs. 5 and 6. 

It is worth paraphrasing these last observations as they address a 
point of key importance. While credit risk models that do {\em not\/} take 
direct economic interactions into account can provide a very reasonable 
fit, when calibrated on historical data which reflect normal economic 
conditions, their predictions would be entirely inadequate when it comes 
to estimating default rates and losses in situations of significant 
economic stress.

Note that the model presented here is suitable for detailed and
comprehensive stress testing, as explicitly demanded within
the regulatory framework of the Basel II accord \cite{BISII}. The
issue of stress testing was addressed in greater detail when the present
model was first introduced in \cite{NeuKuehn04}.

The patterns of economic interactions studied in the present paper
are described by an Erd\"os-R\'enyi random graph. The large
connectivity limit considered in the present investigation further
entails that there is no pronounced heterogeneity in the sets of
economic partners of any one given node. Connectivity distributions
other than Poisson can, however, be handled by suitably adapting the
generating function approach explained in \ref{app:gfa} along the
lines developed in \cite{Hatchett05}, and will be investigated in a
separate publication \cite{Anand+05}. In terms of model fitting
there appear to be a vast number of free parameters in terms of the
interactions $\{J_{ij}\}$ between firms. However, it is important to
realise that to understand the macroscopic behaviour, here only
their low order statistics are relevant.

In the present investigation, we restricted ourselves to analysing
the effects of interactions on default-dynamics and, via default
rates, on loss distributions. More subtle effects such as
credit-quality migration are, as yet, not taken into account, but
could be modelled along similar lines using the dynamics of
interacting multi-state indicator variables. Further assumptions
concerning details of such models would be required, however, and
the full complexity of non-Markovian dynamics would resurface in
such an analysis.

\appendix
\section{Generating Function Analysis}
\label{app:gfa} In this appendix we describe the generating function
approach (GFA) to solve our model, giving  full justification to the
arguments used in section \ref{sec:heuristic}. The reasoning is
relatively standard; we include it here to  make the paper
reasonably self-contained

\subsection{The Generating Function for Correlation Functions}

First we introduce the generating function at fixed value of the macro-economic force
$\eta_0$,
\begin{equation}
Z[\psi|\eta_0] = \left\bra \rme^{-\rmi \sum_{t=0}^{12} \sum_i \psi_{it} n_{it}} \right\ket
\end{equation}
where the angled brackets denote averages over the microscopic dynamics (\ref{eq:micro})
of $n_i$, i.e.
\begin{eqnarray}
Z[\psi|\eta_0] = \sum_{\vn_0,\ldots ,\vn_{12}} P[\vn_0,\ldots,\vn_{12}]\rme^{-\rmi \sum_{t=0}^{12} \sum_i \psi_{it} n_{it}} \ ,
\end{eqnarray}
with $P[\vn_0,\ldots,\vn_{12}]$ denoting the probability of a
sequence of configurations of the entire set of interacting firms
over the risk period of 12 months. The generating function can be
used to compute expectation values and correlation functions via
differentiations with respect to the source fields $\psi_{it}$,
$$
\bra n_{it} \ket = \left . \rmi \frac{\partial Z[\psi|\eta_0]}{\partial \psi_{it}}\right|_{\psi\equiv 0}
\ \ , \qquad
\bra n_{is} n_{jt}\ket = \left . \rmi^2 \frac{\partial^2 Z[\psi|\eta_0]}{\partial \psi_{is} \partial \psi_{jt}}\right|_{\psi\equiv 0}
$$
It is expected that correlation functions averaged over the randomness in the couplings
$J_{ij}$ are dominated by typical realizations of the disorder, hence to describe typical
results an average of the generating function over the disorder,
\be
\overline{Z[\psi|\eta_0]} = \int \prod_{i<j} \rmd P(J_{ij},J_{ji})~Z[\psi|\eta_0]
\ee
is computed.

To proceed, the path-probability $P[\vn_0,\ldots,\vn_{12}]$ at given $\eta_0$ is expressed
in terms of transition probabilities of the Markovian dynamics,
$$
P[\vn_0,\ldots,\vn_{12}]= P(\vn_0)~\prod_{t=0}^{11} P(\vn_{t+1}|\vn_t)\ ,
$$
where
\be
P(\vn_{t+1}|\vn_t)=\prod_i \int \frac{\rmd \xi_{it}}{\sqrt{2\pi}}\rme^{-\frac{1}{2}\sum_i
\xi_{it}^2} \delta_{n_{it+1},f_{it}}
\ee
with
\be
f_{it}=n_{it}+ (1-n_{it})\Theta\left(\sum_j J_{ij} n_{jt}+ \sqrt{1-\rho} ~\xi_{it}
+\sqrt\rho ~\eta_0 - \vartheta_i \right)
\ee
The $\xi_{it}$ integrations appearing in the transition probabilities and the average over
the $J_{ij}$ distribution are facilitated by utilizing $\delta$-distributions to `extract'
the $\xi_{it}$ and the $J_{ij}$ from the Heaviside function in $f_{it}$, using
$$
1= \int \rmd u_{it}~ \delta \Bigg(u_{it} -\sum_j J_{ij} n_{jt}
 - \sqrt{1-\rho} ~\xi_{it} \Bigg) = \int \frac{\rmd u_{it} \rmd \hat u_{it}}{2\pi}~
\rme^{-\rmi\hat u_{it}\left(u_{it}-\sum_j J_{ij} n_{jt} - \sqrt{1-\rho} ~\xi_{it}
\right)}\ .
$$
This gives
\be
\hspace{-1cm} P(\vn_{t+1}|\vn_t)=\int \prod_i \frac{\rmd u_{it} \rmd \hat u_{it}}{2\pi}~
\rme^{\sum_i\left[-\frac{1-\rho}{2}\hat u_{it}^2 -\rmi\hat u_{it}\left(u_{it}
-\sum_j J_{ij} n_{jt} \right)\right]} \prod_i \delta_{n_{it+1},f_{it}}
\ee
with now
\be
f_{it}=n_{it}+ (1-n_{it})\Theta\left(u_{it}+\sqrt\rho ~\eta_0 -\vartheta_i \right)
\label{fit}
\ee
Inserting into the generating function, we get
\bea
Z[\psi|\eta_0] &=& \sum_{\vn_0,\ldots ,\vn_{12}} P(\vn_0)\int \prod_{it} \frac{\rmd
u_{it}  \rmd \hat u_{it}}{2\pi}~\exp\left\{\sum_{it}\Bigg[\frac{1-\rho}{2}\hat (\rmi u_{it})^2
\right .\nn\\
& & \left. -\rmi\hat u_{it}\Bigg(u_{it} -\sum_j J_{ij} n_{jt} \Bigg)
 -\rmi \psi_{it} n_{it}\Bigg]\right\} \prod_{it} \delta_{n_{it+1},f_{it}} \nn
\eea The disorder average affects the $J_{ij}$; it factorises in the
pairs $(i,j)$ and involves the term
$$
\prod_{(i,j)} \overline{D_{ij}}=\prod_{i < j} \overline{
\exp\left\{\rmi\sum_{t} \Big(\hat u_{it}J_{ij} n_{jt}+ \hat u_{jt}
J_{ji} n_{it}\Big)\right\}}^{c,x}
$$
Here the superscripts $c$ and $x$ indicate averages over the
$c_{ij}$ and the $x_{ij}$ in the $J_{ij}$ according to the
statistics (\ref{eq:PJij1})-(\ref{eq:PJij3}). Performing the
$c_{ij}$ average, one gets \bea \prod_{(i,j)} \overline{D_{ij}} &=&
\prod_{i <j} \left\{1+ \frac{c}{N}\overline{\left[\exp\left\{
\Bigg(\frac{J_0}{c} + \frac{J}{\sqrt c} x_{ij}\Bigg) \sum_{t} \rmi
\hat u_{it} n_{jt}
\right .\right .}\right .\nn\\
& & \left .~~~~~~~ \overline{+ \left .\left . \Bigg(\frac{J_0}{c} +
\frac{J}{\sqrt c} x_{ji}\Bigg) \sum_{t} \rmi \hat u_{jt} n_{it}
\right\}-1\right]}^x \right\} \nn \eea The exponential is expanded
using $c\gg 1$. Using (\ref{eq:PJij3}), keeping dominant terms and
re-exponentiating the result one obtains
$$
\prod_{(i,j)} \overline{D_{ij}} \simeq \exp\left\{N\left[J_0\sum_t k_t m_t +
\frac{J^2}{2}\sum_{s,t} \Big[Q_{st}q_{st} + \alpha G_{st}G_{ts}\Big]\right]\right\}
$$
which depends only on the macro-variables
\be
k_t=\frac{1}{N} \sum_i \rmi \hat u_{it}\ \ , \quad m_t =\frac{1}{N} \sum_i  n_{it}
\label{OP1}
\ee
\be
Q_{st}=\frac{1}{N} \sum_i \rmi \hat u_{is} \rmi \hat u_{it}\ ,
\quad q_{st} =\frac{1}{N} \sum_i  n_{is}n_{it} \ ,\quad
G_{st}=\frac{1}{N} \sum_i \rmi\hat u_{is} n_{it}\ .
\label{OP2}
\ee
We thus have
\bea
\fl
\overline{Z[\psi|\eta_0]} =\ \sum_{\vn_0,\ldots ,\vn_{12}} P(\vn_0)\int \prod_{it} \frac{\rmd
u_{it}  \rmd \hat u_{it}}{2\pi}~\exp\left\{\sum_{it}\Bigg[\frac{1-\rho}{2}\hat (\rmi u_{it})^2
-\rmi\hat u_{it} u_{it} -\rmi \psi_{it} n_{it}\Bigg]\right .\nn\\
~~~~~~~~~~ \left. +N\Bigg[J_0\sum_t k_t m_t +
\frac{J^2}{2}\sum_{s,t} \Big[Q_{st}q_{st} + \alpha
G_{st}G_{ts}\Big]\Bigg] \right\} \prod_{it} \delta_{n_{it+1},f_{it}}
\nn \eea Site factorisation in  $\overline{Z[\psi|\eta_0]}$ is
achieved as usual by writing it as an integral over the
macro-variables, using $\delta$-function identities of the form
$$
1 = \int \rmd (Nm_{t})~ \delta \Bigg(N m_{t} -\sum_j  n_{jt} \Bigg)
=\int \frac{\rmd m_{t} \rmd \hat m_{t}}{2\pi/N}~ \rme^{\rmi\hat
m_{t}\left(N m_{t} -\sum_j  n_{jt} \right)}
$$
and analogous ones for the $k_t$, $q_{st}$, $Q_{st}$, and  the
$G_{st}$ to compute densities of state. This results in the
following compact expression for the average generating function \be
\overline{Z[\psi|\eta_0]}=\int {\cal
D}\{\dots\}\exp\left\{N[\Phi+\Psi+\Xi]\right\} \label{pathint} \ee
in which ${\cal D}\{\dots\}$ stands for an differentials of all
order parameters introduced in (\ref{OP1}), (\ref{OP2}) and their
conjugate (hatted) parameters introduced via Fourier-representations
of $\delta$-functions. The functions $\Phi$, $\Psi$, and $\Xi$
appearing in (\ref{pathint}) are given by \bea \Phi &=& J_0\sum_t
k_t m_t + \frac{J^2}{2}\sum_{s,t} \Big[Q_{st}q_{st} +
\alpha G_{st}G_{ts}\Big]\\
\Psi &=& \rmi \sum_t [\hat m_t m_t + \hat k_t k_t] + \rmi \sum_{st} [\hat q_{st} q_{st}
+ \hat Q_{st} Q_{st} + \hat G_{st} G_{st}] \\
\Xi &=& \frac{1}{N}\sum_i \log \sum_{\{n_t\}}\int \prod_t \frac{\rmd \hat u_t \rmd
u_t}{2\pi} \exp\Bigg(-\cS -\rmi\sum_t \psi_{it}n_t\Bigg)~\prod_t
\delta_{n_{t+1},f_{it}}\ .
\label{eq:Xi}
\eea
with $\cS$ denoting the `dynamic action'
\be
\hspace{-2.5cm}
\cS=\sum_t \Bigg[-\frac{1-\rho}{2}(\rmi\hat u_{t})^2 +\rmi\hat u_{t} u_{t} + \rmi\hat m_t n_t
+\rmi \hat k_t \rmi \hat u_t\Bigg]+
\rmi \sum_{st} \Bigg[\hat q_{st}n_s n_t + \hat Q_{st} \rmi\hat u_{s}\rmi\hat u_{t}
+ \hat G_{st} \rmi\hat u_{s} n_{t} \Bigg]
\label{eq:S}
\ee
The third contribution, $\Xi$, in (\ref{pathint}) describes an ensemble of independent
single site dynamical problems.  Thus, to leading order in $N$ we have written our generating
function in terms of an integral which may be computed via a saddle point argument.

\subsection{Saddle Point Problem}
In the saddle-point, variation of our observables gives
\bea
\rmi \hat m_t &=& -J_0 k_t \qquad ~~~~~~\rmi \hat k_t = - J_0 m_t \nn\\
\rmi \hat q_{st} &=& -\frac{J^2}{2}Q_{st} \qquad ~~~~\rmi \hat
Q_{st}=-\frac{J^2}{2}q_{st}   \qquad \rmi \hat G_{st}= -\alpha J^2 G_{ts}
\eea
\bea
m_t &=& \frac{1}{N} \sum_i \bra n_t \ket_{(i)} \qquad
k_t=\frac{1}{N} \sum_i \bra \rmi \hat u_t \ket_{(i)}
\label{mkt}\\
q_{st}&=&\frac{1}{N} \sum_i \bra n_s n_t \ket_{(i)}  \quad
Q_{st}= \frac{1}{N} \sum_i \bra \rmi\hat u_s \rmi\hat u_t \ket_{(i)}\\
G_{st}&=&\frac{1}{N} \sum_i \bra \rmi\hat u_s n_t \ket_{(i)}
\label{qQG} \eea with $\bra \dots\ket_{(i)}$ denoting averages
evaluated wrt effective single site dynamics at $i$. \be \bra
\dots\ket_{(i)}= \frac{\sum_{\{n_t\}}\int \prod_t \frac{\rmd \hat
u_t \rmd u_t}{2\pi} (\dots)\exp\Big(-\cS \Big)\prod_t
\delta_{n_{t+1},f_{it}}}{\sum_{\{n_t\}}\int \prod_t \frac{\rmd \hat
u_t \rmd u_t}{2\pi} \exp\Big(-\cS \Big)\prod_t
\delta_{n_{t+1},f_{it}}} \label{i-av} \ee In the usual manner
\cite{Dominicis78} averages involving conjugate fields $\rmi \hat
u_t$ describe response functions, i.e. perturbations of expectation
values wrt external fields, so that averages involving nothing but
conjugate variables correspond to perturbations of a constant and
will therefore vanish. Moreover, causality implies that $G_{st}$,
which describes the response of the average fraction of defaulted
companies to at time $t$ to perturbations at time $s$ must vanish
for $s\ge t$. At the saddle point, therefore, we have $k_t\equiv 0$,
$\rmi\hat m_t \equiv 0$, $Q_{st} \equiv 0$, $\rmi \hat q_{st} \equiv
0$, and $G_{st} = 0$ for  $s\ge t$, thus $\rmi \hat G_{st}= 0$ for
$s\le t$.

With these observations we find that the functions $\Phi$ and $\Psi$ appearing
in the average generating function (\ref{pathint}) are zero at the saddle point,
\be
\Phi = 0 \ , \qquad \Psi = 0\ ,
\ee
and the dynamic action $\cS$ of (\ref{eq:S}) simplifies to
\be
\hspace{-2cm}
\cS = -\frac{1}{2} \sum_{st} \Big[(1-\rho)\delta_{st} + J^2 q_{st}\Big]
 \rmi\hat u_{s}\rmi\hat u_{t} + \sum_t \rmi\hat u_{t} \Bigg (u_{t}
-J_0 m_t -\alpha J^2 \sum_{s<t} G_{st}n_{s}\Bigg) \ee With this form
of the dynamic action the system-dynamics is described by an
ensemble independent effective single-node stochastic process of the
form \be \hspace{-1cm} n_{t+1}=f_{\vartheta t} \equiv n_{t}
+(1-n_{t})\Theta\Bigg(J_0 m_t+\alpha J^2 \sum_{s<t} G_{st}n_{s}
+\sqrt\rho \eta_0- \vartheta +\phi_t\Bigg) \label{eq:sspin1} \ee the
details of which are self-consistently specified by macroscopic
properties of the system via the saddle point equations, in that
each single site process (i) depends on the dynamics of the
macroscopic fraction of defaulted nodes $m_t$, (ii) the original
Gaussian white noise is replaced by a coloured Gaussian noise
$\phi_t$ with correlations depending on $q_{st}$
$$
\bra \phi_t\ket= 0 ~~,~~ \bra \phi_s\phi_t\ket= (1-\rho) \delta_{st} + J^2 q_{st} \ ,
$$
and (iii) a memory term appears in the dynamics, if there is some
degree of symmetry in the interactions, i.e. if  $\alpha\ne 0$.

The only site-dependence in the averages $\bra \dots\ket_{(i)}$
appearing the fixed point equations (\ref{mkt}) - (\ref{qQG}) comes
from the $\vartheta_i$ dependence in the update rules $f_{it}$. By
the law of large numbers, the sums can therefore be evaluated as an
average over the $\vartheta$-distribution in the large $N$ limit
$$
\frac{1}{N}\sum_i \bra \dots\ket_{(i)} \longrightarrow \int \rmd \vartheta p(\vartheta)
\bra \dots\ket_{(\vartheta)} \equiv
\left\bra\bra \dots\ket_{(\vartheta)}\right\ket_\vartheta
$$
in which $\bra \dots\ket_{(\vartheta)}$ has the same structure as
(\ref{i-av}), except for the fact that  the dynamical constraints
$f_{it}$ of (\ref{fit}) are replaced by the $f_{\vartheta t}$ of
(\ref{eq:sspin1}). The saddle point equations thus take the form
$$
m_t = \left\bra\bra n_t\ket_{(\vartheta)}\right\ket_\vartheta ~~,~~
q_{st} = \left\bra\bra n_s n_t \ket_{(\vartheta)}\right\ket_\vartheta ~~,~~ G_{st} = \left\bra \bra \rmi\hat u_s n_t \ket_{(\vartheta)}\right\ket_\vartheta
$$

\subsection{Simplification of the Single Node Equation}
The single node equation (\ref{eq:sspin1}) is complicated by the fact
that it is non-Markovian, containing a correlation function coupled to
the noise term $q_{st}$ and a retarded self-interaction $G_{st}$. This
latter term encodes the physics that a firms performance at time $t$
is influence by its neighbours, themselves dependent on the firm
itself at times $s < t$, via loops in our network of corporate
interactions --- in particular short loops arising through correlated
bi-directional interactions. However, as we argued in section
\ref{sec:heuristic}, if a firm is bankrupt at time $s$ then the performance
of partner firms at time $t$ is irrelevant, since the firm will still be
bankrupt. In the alternative case when the firm is solvent at time $t$, it is
clear from the definitions in the dynamics that it must have been
solvent at time $s < t$ and thus cannot have affected its partner
terms at that time. Thus, the retarded self-interaction is zero.

There is a second simplifying feature in (\ref{eq:sspin1}) related to
the statistics of the coloured noise within our system. On multiplying
(\ref{eq:sspin1}) on both sides by $n_s$ with $s<t$ and and first
averaging over the noise $\bphi_t=(\phi_1,\phi_2,\dots,\phi_t)$ one
finds $\bra n_s n_t \ket_{(\vartheta)} = \bra n_{{\rm min}(s,t)}
\ket_{(\vartheta)}$ at fixed $\vartheta$, since if $n_{s} = 1$, then
$n_{t} = 1$ due to the absorbing nature of the defaulted state, whereas if
$n_{s} = 0$,  so is the product $n_s n_t$, irrespectively of $n_{t}$.
As a consequence we have $q_{st} = m_{{\rm min}(s,t)}$, and thus
\begin{equation}
\bra \phi_s \phi_t \ket = (1-\rho) \delta_{st} + J^2 m_s\ , \qquad  s \leq t\ ,
\end{equation}
Having seen that the memory term in the dynamics vanishes, it transpires that
only the equal-time version of the noise correlation $\bra \phi_t \phi_t \ket
= 1-\rho + J^2 m_t$ is required to propagate the order parameter $m_t$.
One needs
$$
m_{t+1} = \left\bra\bra n_{t+1}\ket_{(\vartheta)}\right\ket_\vartheta
=m_t + \left\bra \Big\bra(1-n_t) \Theta(J_0m_t + \sqrt{\rho}\eta_0 -\vartheta + \phi_t)\Big\ket_{(\vartheta)} \right\ket_\vartheta
$$
In order to evaluate the average $\bra \dots \ket_{(\vartheta)}$ over the correlated
noise in the second term, convert the probability density $p(\bphi_t)= p(\phi_1,
\phi_2,\dots,\phi_t)$ into $p(n_t,\phi_t)$ --- the joint probability density that
 the node-variable takes  value $n_t$  ($n_t=0$ or $n_t=1$)  and the noise variable
at time $t$ is in an infinitesimal interval around $\phi_t$; formally one can write
this as
$$
p(n_t,\phi_t)= \int \rmd \bphi_{t-1} p(\bphi_{t})
\delta_{n_t,n_t(\vartheta,{\small\bphi_{t-1}})}
 = \int \rmd \bphi_{t-1} p(\bphi_{t-1},\phi_t)
\delta_{n_t,n_t(\vartheta,{\small\bphi_{t-1}})}
$$
where $n_t(\vartheta,\bphi_{t-1})$ is the value of $n_t$ for the
specific $\vartheta$ under consideration, and a given noise history
$\bphi_{t-1}$. Writing the joint probability in terms of a
conditional as
$$
p(n_t,\phi_t)= p(n_t|\phi_t) p(\phi_t)\ ,
$$
and noting that  $p(n_t|\phi_t)$ must be independent of the conditioning by causality,
and finally using
$$
p(n_t) = \bra n_t \ket_{(\vartheta)} \delta_{n_t,1} + \Big(1- \bra n_t \ket_{(\vartheta)} \Big) \delta_{n_t,0}
$$
for a given $\vartheta$ one finally obtains
\be
\hspace{-2cm}
m_{t+1} =\left\bra\bra n_{t+1}\ket_{(\vartheta)}\right\ket_\vartheta
= m_t + \left\bra \frac{1- \bra n_t \ket_{(\vartheta)}}{2}\left[1+
\erf\left(\frac{J_0 m_t + \sqrt\rho ~\eta_{0}-\vartheta}
{\sqrt{2(1-\rho + J^2 m_t)}}\right)\right]\right\ket_{\vartheta}
\ee
which agrees with the result of our heuristic reasoning in Sec \ref{sec:heuristic}.
Note that the condition $c/N \to 0$ is not needed in the present argument.\\

{\bf Acknowledgements} It is a pleasure to thank Peter Neu for valuable
discussions on a range of problems dealt with in the present contribution.\\



\begin{thebibliography}{99}

\bibitem{Keenan00}
S. Keenan, {\em Historical default rates of corporate bond issues,
1920-2000}, Moody's Investor Services, (2000)

\bibitem{BISII}
Basel~Committee on~Banking~Supervision, {\em International Convergence
of Capital Measurement and Capital Standards A Revised Framework},
URL: {\tt http://www.bis.org}, Basel (Nov 2005).

\bibitem{CM}
J.P. Morgan Global Research, {\em CreditMetrics$^{\rm TM}$: The
Benchmark for Understanding Credit Risk}, Technical Document,
{\tt URL: www.creditmetrics.com}, New York (1997).

\bibitem{CRisk}
Credit Suisse First Boston, {\em Credit Risk+: A Credit Risk
Management Framework}, Technical Document, {\tt URL:
www.csfb.com/creditrisk}, New York (1997).

\bibitem{Kealhofer98}
S. Kealhofer, {\em Portfolio Management of Default Risk}, Net
Exposure, Vol. {\bf 1} (2), (1998); KMV Corporation, {\em Credit
Monitor$^{\rm TM}$ and Portfolio Manager$^{\rm TM}$}, New York,
(1996), {\tt URL: www.KMV.com};  O. A. Vasicek, {\em The loan Loss
Distribution}, Technical Report, KMV Corporation (1997).

\bibitem{Davis00}
M. Davis and V. Lo, {\em Infectious Defaults}, Quantitative Finance
{\bf 1}, p. 382-387 (2001); {\em Modelling Default Correlation in
Bond Portfolios}, in Mastering Risk Volume 2: Applications, ed.
Carol Alexander, Financial Times Prentice Hall, p. 141-151 (2001);
M. Crowder, M. Davis and G. Giampieri, {\em A Hidden Markov
 Model of default interaction}, Quantitative Finance {\bf 5}, p.
 27-34, (2005).

\bibitem{Lando98}
D. Lando {\em On Cox processes and credit risky securities}, Rev.
Derivatives Res. {\bf 2}, p. 99-120, (1998)

\bibitem{Duffie98}
D. Duffie and K. Singleton {\em Modelling term structures of
defaultable bonds} Rev. Financial Stud. {\bf 12}, p. 687-720, (1999)

\bibitem{Jarrow01}
R. Jarrow and F. Yu, {\em Counter-party Risk and the Pricing of
Defaultable Securities}, Journal of Finance {\bf 56}, p. 1765
(2001); F. Yu, {\em Correlated Defaults in Intensity Based Models}
Mathematical Finance (to appear).

\bibitem{Rogge02}
E. Rogge and P. Sch\"onbucher, {\em Modelling Dynamic Portfolio
Credit Risk}, Working Paper, Imperial College, London, ABN Amro
Bank, London, and ETH, Z\"urich (February 2003); P. Sch\"onbucher
and D. Schubert {\em Copula-dependent Default Risk in Intensity
Models} Working Paper, Department of Statistics, Bonn University,
(2001).

\bibitem{Jarrow03}
F. Yu, R. Jarrow and D. Lando, {\em Default Risk and
Diversification: Theory and Empirical Implications} Mathematical
Finance {\bf 15}(1), p. 1-26, (2005).

\bibitem{Das05}
S. R. Das, D. Duffie and N. Kapadia, {\em Common Failings: How
Corporate
 Defaults are Correlated}, Working Paper, Santa Clara University,
 Stanford University, University of Massachusetts, (February 2005).

\bibitem{Yu03}
F. Yu, {\em Default Correlation in Reduced-Form Models}, Journal of
Investment Management, {\bf 3}(4), p. 33-42, (2005).

\bibitem{Duffie99} D. Duffie and K. Singleton {\em Simulating
Correlated Defaults} Working Paper, Graduate School of Business,
Stanford University, (May 1999)

\bibitem{Giesecke03}
K. Giesecke, {\em A simple exponential model for dependent
defaults},  Journal of Fixed Income, {\bf 13}(3), 74-83, (2003); K.
Giesecke , {\em Successive Correlated Defaults: Pricing Trends and
Simulations }, Computing in Economics and Finance No 247, Society
for Computational Economics, (2003).

\bibitem{Weber02}
K. Giesecke and S. Weber, {\em Credit Contagion and Aggregate
Losses}, Journal of Economic Dynamics and Control (to appear), K.
Giesecke and S. Weber, {\em Cyclical Correlation, Credit Contagion
and Portfolio Losses},Journal of Banking and Finance {\bf 28}(12),
p. 3009-3036, (2004).

\bibitem{Tasche03}
D. Tasche and U. Theiler, {\em Calculating Concentration-Sensitive
Capital with Conditional Value-at-Risk}, in Operations Research
2003, D. Ahr, R. Fahrion, M. Oswald, G. Reinelt (eds.) Springer, p.
261-268, (2004).

\bibitem{Eglo03}
D. Egloff, M. Leippold, and P. Vanini, {\em A Simple Model of Credit
Contagion}, EFA 2004 Maastricht Meetings Paper No. 1142, (January 2,
2004).

\bibitem{Gordy00}
M. B. Gordy, {\em A Comparative Anatomy of Credit Risk Models},
Journal of Banking and Finance, Vol. {\bf 24}, p. 119-149 (2000).

\bibitem{Gordy01}
M. B. Gordy, {\em A Risk-Factor Model Foundation for Ratings-Based
Capital Rules}, Journal of Financial Intermediation , Vol. {\bf
12}(3), p. 199-232 (2002).

\bibitem{NeuKuehn04}
P. Neu and R. K\"uhn, {\em Credit risk Enhancement in a Network of Interdependent
Firms}, Physica  A {\bf 342}, 639--655 (2004)

\bibitem{Merton74}
R. Merton, {\em On the Pricing of Corporate Debt: The Risk Structure
of Interest Rates}, Journal of Finance, Vol. {\bf 29}, p. 449-470
(1974).

\bibitem{Algo99}
Ian Iscoe, Alex Kreinin and Dan Rosen, {\em An Integrated Market and
Credit Risk Portfolio Model}, Algorithmics Research Quarterly, Vol.
{\bf 2} (3), p. 21-37 (September 1999).

\bibitem{Moodys97}
Moody's Investment Services, {\em The Binominal Expansion
Technique}, {\tt URL: www.moodys.com} (1997).

\bibitem{KuehnNeu03}
R. K\"uhn and P. Neu, {\em Functional Correlation Approach to
Operational Risk in Banking Organizations},  Physica A {\bf 322},
650--666 (2003).

\bibitem{Bollobas01}
B. Bollob\`as, {\em Random Graphs}, (Cambridge Univ. Press, Cambridge, 2001)

\bibitem{Barabasi99}
A.L. Barab\'asi and R. Albert, {\em Emergence of Scaling in Random
Networks}, Science, {\bf 286}, p. 509-512, (1999); D.J. Watts and
S.H. Strogarz {Collective dynamics of 'small-world' networks}
Nature, {\bf 393}, p. 440-442 (1998).

\bibitem{Per+04}
J.P.L. Hatchett, B. Wemmenhove, I P\'erez Castillo, T.
Nikoletopoulos, N.S. Skantzos and A.C.C. Coolen, {\em Parallel
dynamics of disordered Ising spin systems on finitely connected
random graphs} J. Phys. A: Math. Gen. {\bf 37}, 6201-6220 (2004); I
P\'erez Castillo, B. Wemmenhove, J.P.L. Hatchett, A.C.C. Coolen,
N.S. Skantzos and T. Nikoletopoulos, {\em Analytic Solution of
Attractor Neural Networks on Scale-Free Graphs}, J. Phys. A: Math.
Gen.  {\bf 37}, 8789--8799 (2004); T. Nikoletopoulos, A.C.C. Coolen,
I P\'erez Castillo, N.S. Skantzos, J.P.L. Hatchett and B. Wemmenhove
{\em Replicated transfer matrix analysis of Ising spin models on
'small world' lattices} J. Phys. A:Math. Gen., {\bf 37}, p.
6455-6475, (2004).

\bibitem{Hatchett05}
J.P.L. Hatchett, {\em  Parallel dynamics of disordered Ising spin
systems on random graphs}, AIP Conf. Proc. {\bf 776}, 150, (2005);

\bibitem{Dominicis78}
C.~De Dominicis,  {\em Dynamics as Substitute for Replicas in
Systems with Quenched Random Impurities}, Phys. Rev. B {\bf 18}
4913-4919 (1978)

\bibitem{Derrida87}
B. Derrida, E. Gardner, A. Zippelius, {\em An exactly solvable
asymmetric neural network model} Europhys. Lett. 4, 167-173 (1987);
R. Kree and A. Zippelius {\em Asymmetrically diluted neural
networks} in E. Domany, J. L. van Hemmen, and K. Schulten, K eds,
Models of Neural Networks. Springer, Berlin (1991); J.P.L. Hatchett
and A. C. C. Coolen {\em Asymmetrically extremely dilute neural
networks with Langevin dynamics and unconventional results} J. Phys.
A: Math. Gen. {\bf 37}, p. 7199-7212, (2004); T.L.H. Watkin and D.
Sherrington {\em  A neural network with low symmetric connectivity }
Europhys.Lett. {\bf 14}, p. 791-796, (1991); T.L.H. Watkin D. and
Sherrington  {\em The parallel dynamics of a dilute symmetric
Hebb-rule network} J. Phys. A:Math. Gen. {\bf 24}, p. 5427-5433,
(1991).

\bibitem{Phasetrans} L.E. Reichl, {\em A Modern Course in
Statistical Physics}, Wiley-VCH (1998).

\bibitem{Anand+05}
K. Anand, J.P.L. Hatchett, and R. K\"uhn, work in progress (2005)

\end{thebibliography}
\end{document}